\documentclass{PoS}
\usepackage[utf8]{inputenc}
\usepackage{xcolor}
\usepackage{epsfig}
\usepackage{amsmath}
\usepackage{xfrac}
\usepackage{tikz}
\usetikzlibrary{decorations.pathreplacing,calc}
\usetikzlibrary{shapes.misc, positioning}
\usetikzlibrary{fadings}

\newcommand{\s}[1]{\texttt{#1}}
\newcommand{\tikzmark}[1]{\tikz[overlay,remember picture] \node (#1) {};}

\title{Identified kaon production in Ar+Sc collisions at SPS energies}
\ShortTitle{Identified kaon production in Ar+Sc collisions at SPS energies}

\author{\speaker{Maciej P. Lewicki}\\
for the NA61/SHINE Collaboration\\
University of Wrocław\\
E-mail: \email{maciej.piotr.lewicki@cern.ch}}

\abstract{NA61/SHINE is a fixed target experiment at the CERN Super Proton Synchrotron. The main goals of the experiment are to discover the critical point of strongly interacting matter and to study the properties of the onset of deconfinement. In order to reach these goals, a study of hadron production properties is performed in nucleus-nucleus, proton-proton and proton-nucleus interactions as a function of collision energy and size of the colliding nuclei. In this talk, the newest preliminary results on kaon spectra produced in \s{Ar+Sc} collisions at three beam momenta (30\textit{A}, 40\textit{A} and 75\textit{A}) will be shown. The distributions of transverse mass and rapidity will be compared with results of NA61/SHINE (\s{p+p}, \s{Be+Be}) and NA49 (\s{Pb+Pb}, \s{C+C}, \s{Si+Si}), as well as with available world data.}

\FullConference{Critical Point and Onset of Deconfinement\\
7-11 August, 2017\\
The Wang Center, Stony Brook University, Stony Brook, NY}

\begin{document}

\section{Introduction}
NA61/SHINE is a fixed target spectrometer~\cite{facility} located in CERN's North Area, utilizing the SPS proton, ion and hadron beams. Tracking capabilities are provided by four large volume Time Projection Chambers (TPC), two of which are located in magnetic fields. The Projectile Spectator Detector (PSD), a zero degree, modular calorimeter, is used to determine the centrality of the collisions.

The aim of the experiment is to explore the QCD phase diagram $(\mu_B,T)$ by a two-dimensional scan in collision energy and system size. The yields of hadrons produced in the collisions are studied for indications of the onset of deconfinement (\textit{kink}, \textit{horn} and \textit{step}) and the critical point of the phase transition (hill in event-by-event fluctuations). This paper discusses the analysis of \s{Ar+Sc} collisions, the next step of the system size scan, following the earlier analysis of \s{p+p} and \s{Be+Be} reactions. Comparison of obtained kaon spectra gives important insight into the dynamics of ion collisions in the transition region between reactions of light and heavy nuclei.


\section{$dE/dx$ Method}
Charged particle identification in the NA61/SHINE experiment is based on the measurement of the ionization energy loss $dE/dx$ in the gas of the TPCs and of the time of flight $tof$ obtained from the ToF-L and ToF-R walls. In the region of the relativistic rise of the ionization at large momenta the measurement of $dE/d$x alone allows particle identification. The acceptance region of this identification method can be seen in Fig.~\ref{fig:ptvy} for each employed beam momentum.

Time projection chambers provide measurements of energy loss $dE/dx$ of charged particles in the chamber gas along their trajectories. Simultaneous measurements of $dE/dx$ and $p_{lab}$ allow for distinction between particle species. Here $dE/dx$ is calculated as the truncated mean (smallest 50\%) of cluster charges measured along the track trajectory.

The contributions of $e^+, e^-, \pi^+, \pi^-, K^+, K^-, p, \bar{p} \textrm{~and~} d$ are obtained by fitting the $dE/dx$ distributions separately for positively and negatively charged particles in bins of laboratory momentum $p_{lab}$ and transverse momentum $p_T$ (see Ref.~\cite{SZpp} for details). In order to ensure similar particle multiplicities in each bin, 13 logarithmic bins are chosen in $p_{lab}$ in the range 5--100 GeV/\textit{c} to cover the full acceptance of the $dE/dx$ method. Furthermore, the data are binned in 20 equal $p_T$ intervals in the range 0--2 GeV/\textit{c}.

\begin{figure}[h]
\centering{
\includegraphics[width=0.45\textwidth]{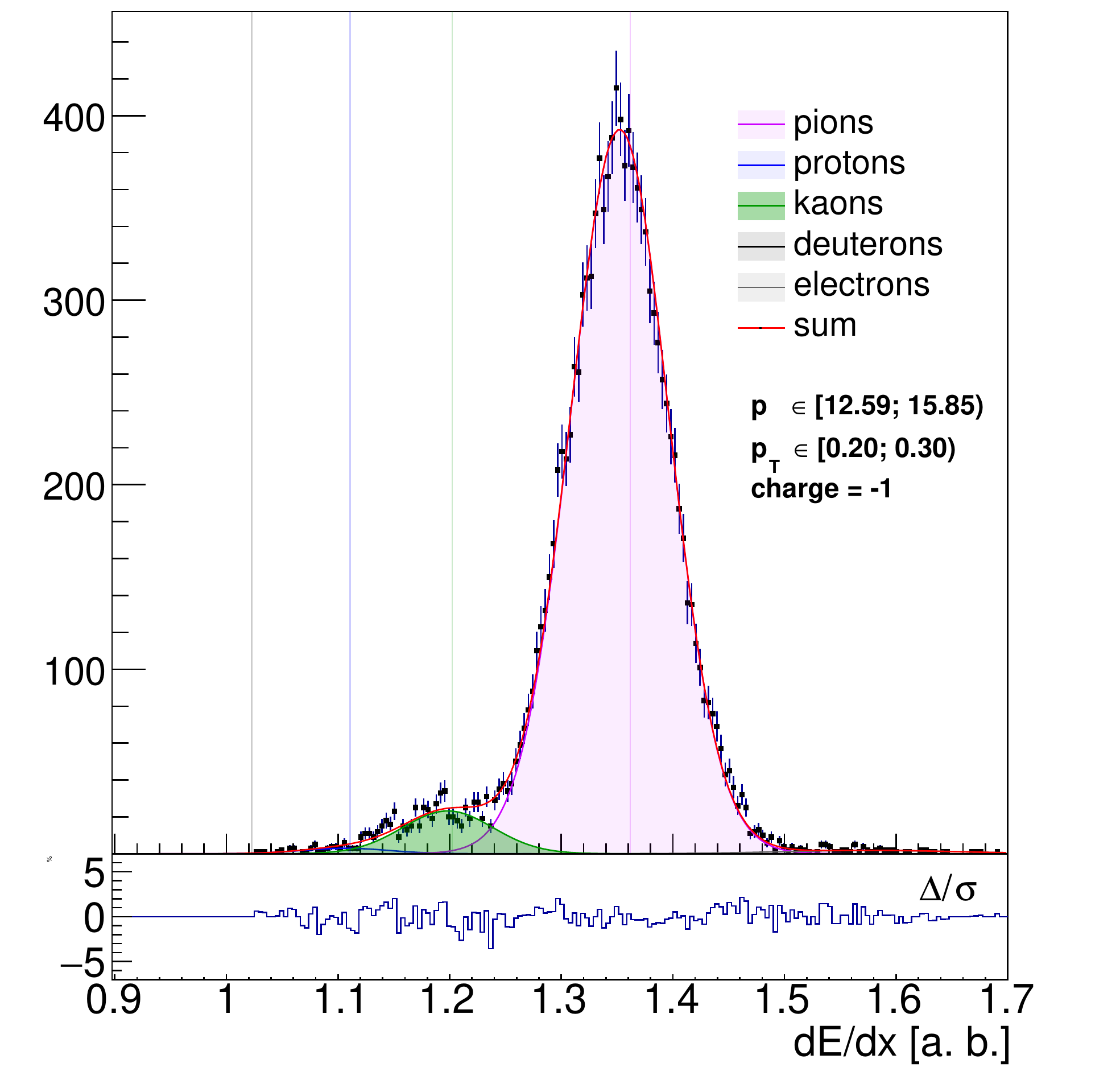}
\includegraphics[width=0.45\textwidth]{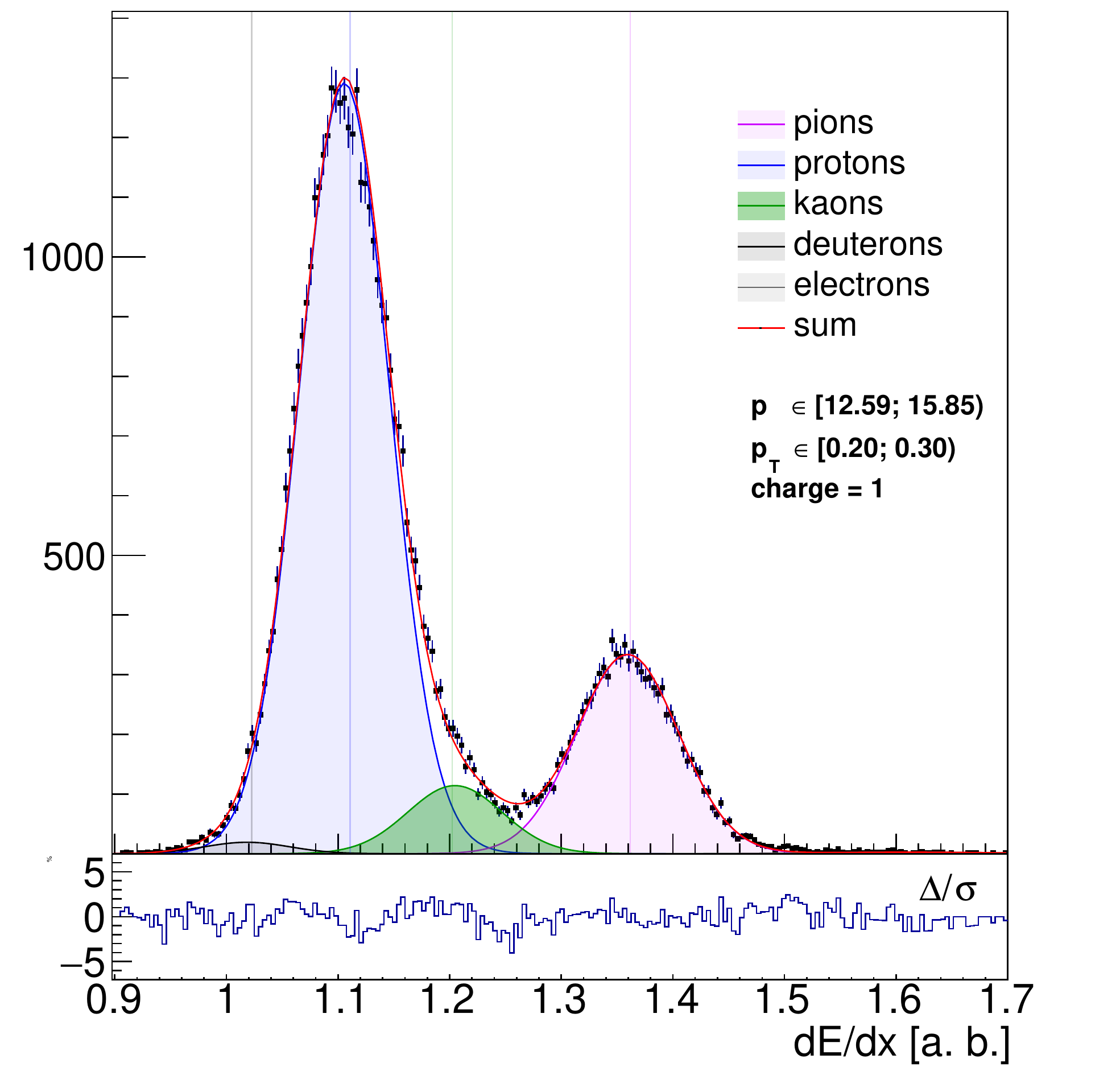}
}
\caption{Example distributions of $dE/dx$ and particle yield fits (residuals shown in lower histograms) in a single bin of the 30\textit{A} GeV/\textit{c} data. Fits were performed in 13 logarithmic bins in $p\in~$[5,~100] GeV/\textit{c} and 20 linear bins in $p_T \in$~[0.0,~2.0] GeV/\textit{c}.}
\label{fig:fits}
\end{figure}

Fits to the $dE/dx$ distributions in these intervals (see examples in Fig.~\ref{fig:fits}) consider five particle types ($i = \pi^{\pm},~K^{\pm},~p/\bar{p},~e^{\pm},~d$). The signal shape for a given particle type is parametrized as the sum of asymmetric Gaussians with widths $\sigma_{i, l}$ depending on the particle type $i$ and the number of points $l$ measured in the TPCs. Simplifying the notation in the fit formulas, the peak position of the $dE/dx$ distribution for particle type $i$ is denoted as $x_i$ . The contribution of a reconstructed particle track to the fit function reads:
\begin{equation}
f(x)~~=~\sum_{i=p,K,\pi,e,d} N_i \frac{1}{\sum_l n_l} ~\sum_l \frac{n_l}{\sqrt{2\pi}~\sigma_{i,l}} ~\exp\left[-\frac{1}{2}\left( \frac{x-x_i}{(1 \pm \delta)~\sigma_{i,l}} \right)^2\right]
\label{eq:fits}
\end{equation}
where $x$ is the $dE/dx$ of the particle, $n_l$ is the number of tracks with number of points $l$ in the sample and $N_i$ is the amplitude of the contribution of particles of type $i$. The second sum is the weighted average of the line-shapes from the different numbers of measured points (proportional to track-length) in the sample. The quantity $\sigma_{i,l}$ is written as:
\begin{equation}
\sigma_{i,l} = \frac{\sigma_0}{\sqrt{l}} \left( \frac{x_i}{x_{\pi}} \right)^{\alpha}
\label{eq:sigma}
\end{equation}
where $\sigma_0$ is assumed to be common for all particle types and $\alpha = 0.625$ is a universal constant. 

\subsection{Probability method}
\begin{figure}[h]
\centering{\includegraphics[width=0.9\textwidth]{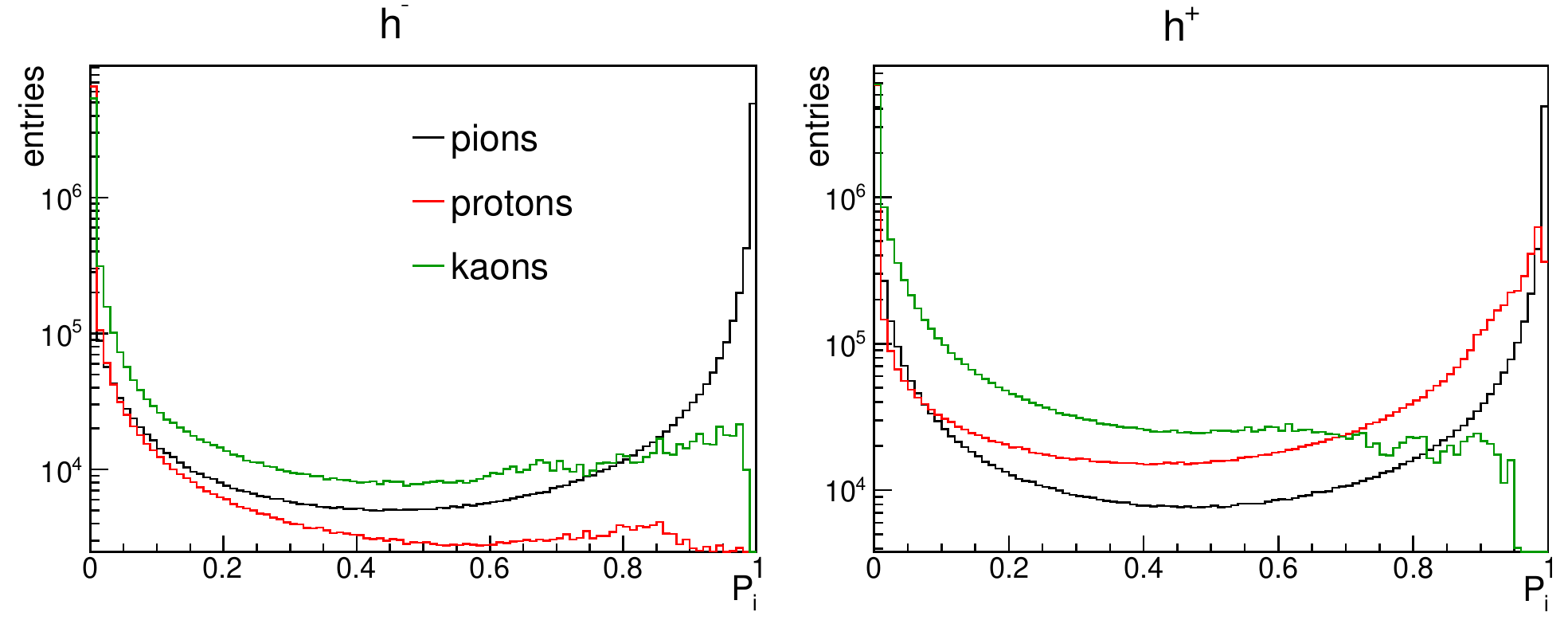}}
\caption{Distribution of particle identification probabilities, obtained for beam momentum of 75\textit{A} GeV/\textit{c}. Left: negatively charged hadrons ($h^-$), right: positively charged hadrons ($h^+$).}
\label{fig:prob}
\end{figure}

The probability method (proposed in Ref.~\cite{MGid}) is used to transform particle spectra obtained in $(p_{lab}, p_T)$ bins to spectra in $(y, p_T)$ bins. The fit results allow to calculate the probability $P_i$ that a measured particle is of the type $i = \pi^{\pm}, K^{\pm}, p, \bar{p}, e^{\pm}, d$:
\begin{equation}
P_i\left(p,p_T,\sfrac{dE}{dx}\right)~~=~~\dfrac{f_i\left(p,p_T,\sfrac{dE}{dx}\right)}{\sum_i f_i \left(p,p_T,\sfrac{dE}{dx}\right)}
\label{eq:prob}
\end{equation}
As an example, Fig.~\ref{fig:prob} shows the distribution of probabilities $P_i$ for being a $ \pi^{\pm}, K^{\pm}, p, \bar{p}$ for all particles from the 75\textit{A}~ GeV/\textit{c} data. The number $n_i$ of particles of type~$i$~in a given kinematical bin $(y, p_T)$ can then be calculated:
\begin{equation}
n_{i~\in~\{\pi^{\pm}, K^{\pm}, p, \bar{p}, e^{\pm}, d\}}~~=~~ \sum_{j=1}^{m} P_i
\label{eq:yields}
\end{equation}
where the summation runs over the number of particles $m$ in the bin.

\section{Data analysis}

\subsection{Event selection}
\begin{figure}[h]
\centering{\includegraphics[width=0.98\textwidth]{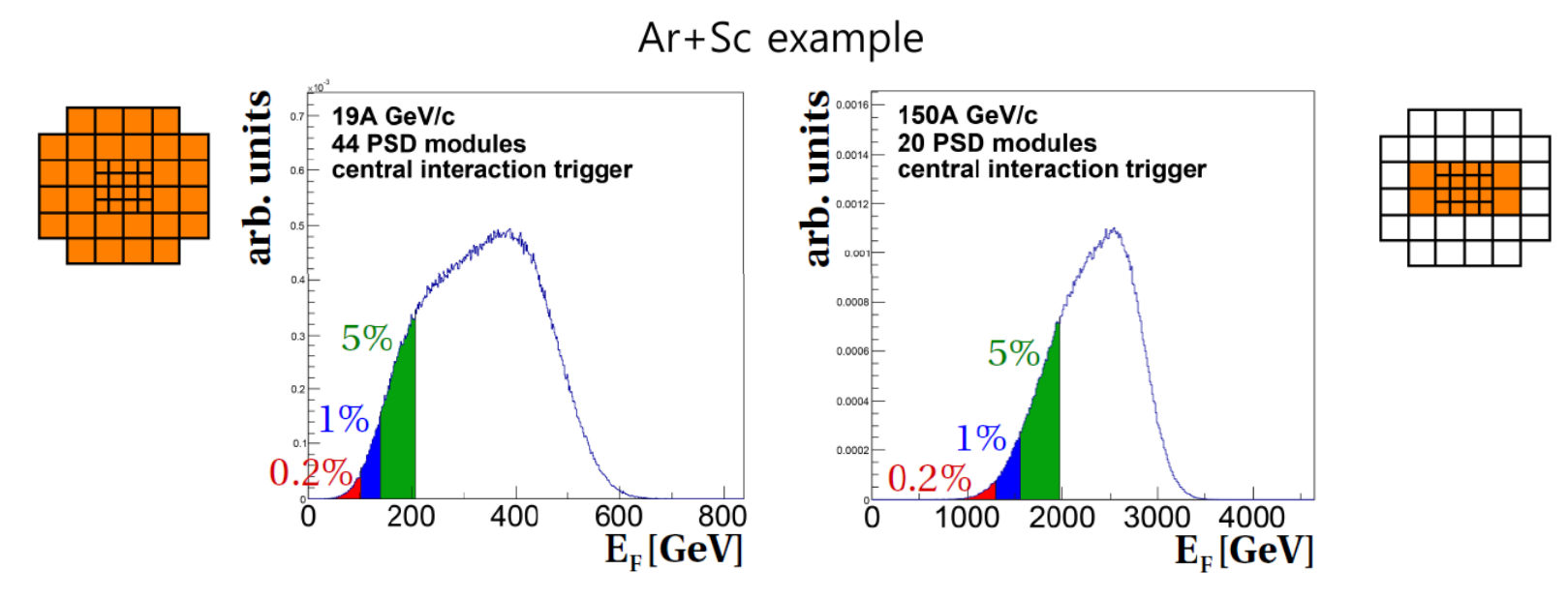}}
\caption{Event "violence" ($\approx$centrality) selection using the forward energy ($E_F$) measured by the PSD calorimeter. Distributions of $E_F$ are shown for 19\textit{A} (left) and 150\textit{A} GeV/\textit{c} (right) beam momentum. Orange shaded areas on the front face diagram of the PSD indicate modules selected for calculation of $E_F$. Choice of modules is based on correlation studies between $E_F$ and track multiplicity measured by the TPCs.}
\label{fig:centr}
\end{figure}

The events recorded by the NA61/SHINE spectrometer were selected for a well reconstructed interaction vertex in the target and the "violence" ($\approx$centrality) of the collisions. Event centrality classes were determined using the PSD calorimeter, located most downstream on the beam line. It measures predominantly the energy $E_F$ carried by projectile spectators, the non-interacting nucleons of the beam nucleus. The distribution of $E_F$ was used to define and select event classes corresponding to collision centrality intervals (see Fig.~\ref{fig:centr}).

\subsection{Corrections and errors}
Corrections of the raw data were based on simulation of \s{Ar+Sc} interactions using the EPOS-1.99~\cite{EPOS} (version CRMC 1.5.3.) model and the GEANT-3.2 code for particle transport and detector simulation (see Ref.~\cite{pp}). Centrality classes in the model calculations were selected by the number of forward spectator nucleons.

Results presented in the plots are shown with statistical uncertainties only. These come from two sources: the experimental data and the simulation-based corrections. The contribution of the latter is insignificant ($<0.1$\%).

Based on the previous analysis of \s{Be+Be}~\cite{BeBe} and \s{p+p}~\cite{SZpp, pp} reactions, systematic errors were estimated at a level of 5\%-10\%.

\section{Transverse Momentum Spectra}
\begin{figure}[h]
	\footnotesize
\centering{30\textit{A} GeV/\textit{c}\hspace{0.35\textwidth}40\textit{A} GeV/\textit{c}}\\
\centering{\includegraphics[width=0.45\textwidth]{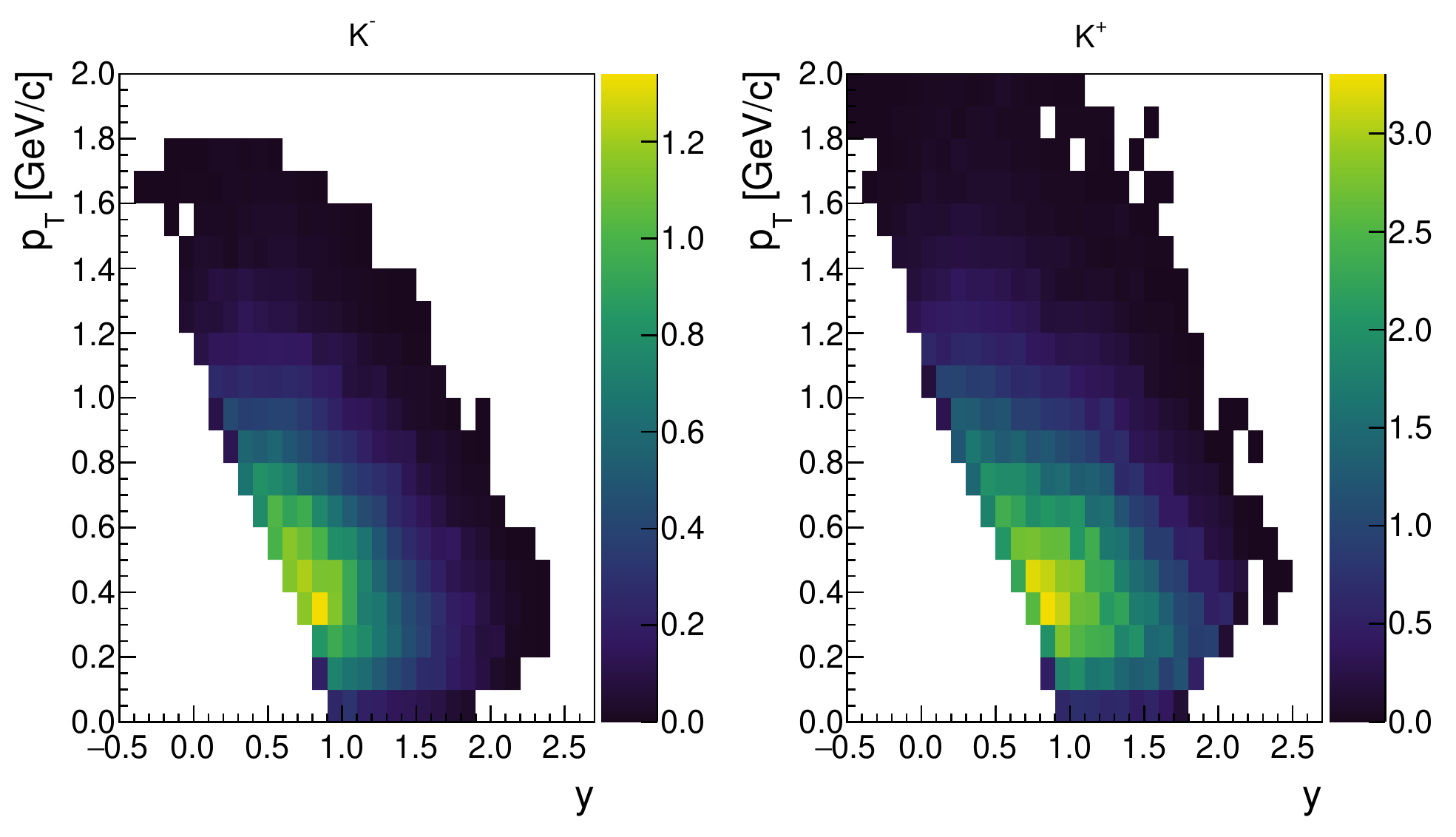}
\includegraphics[width=0.45\textwidth]{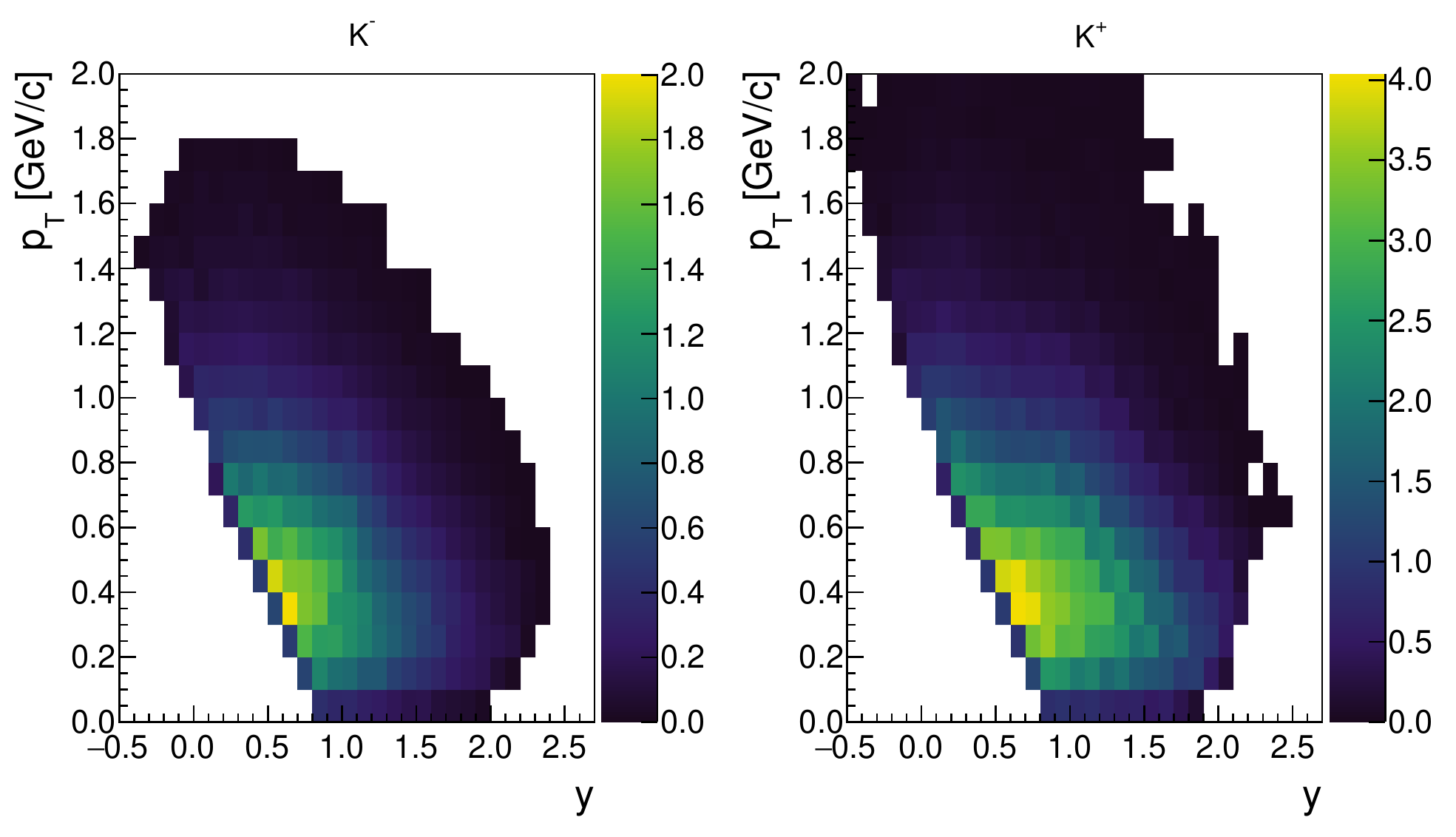}}\\
\centering{75\textit{A} GeV/\textit{c}}\\
\centering{\includegraphics[width=0.45\textwidth]{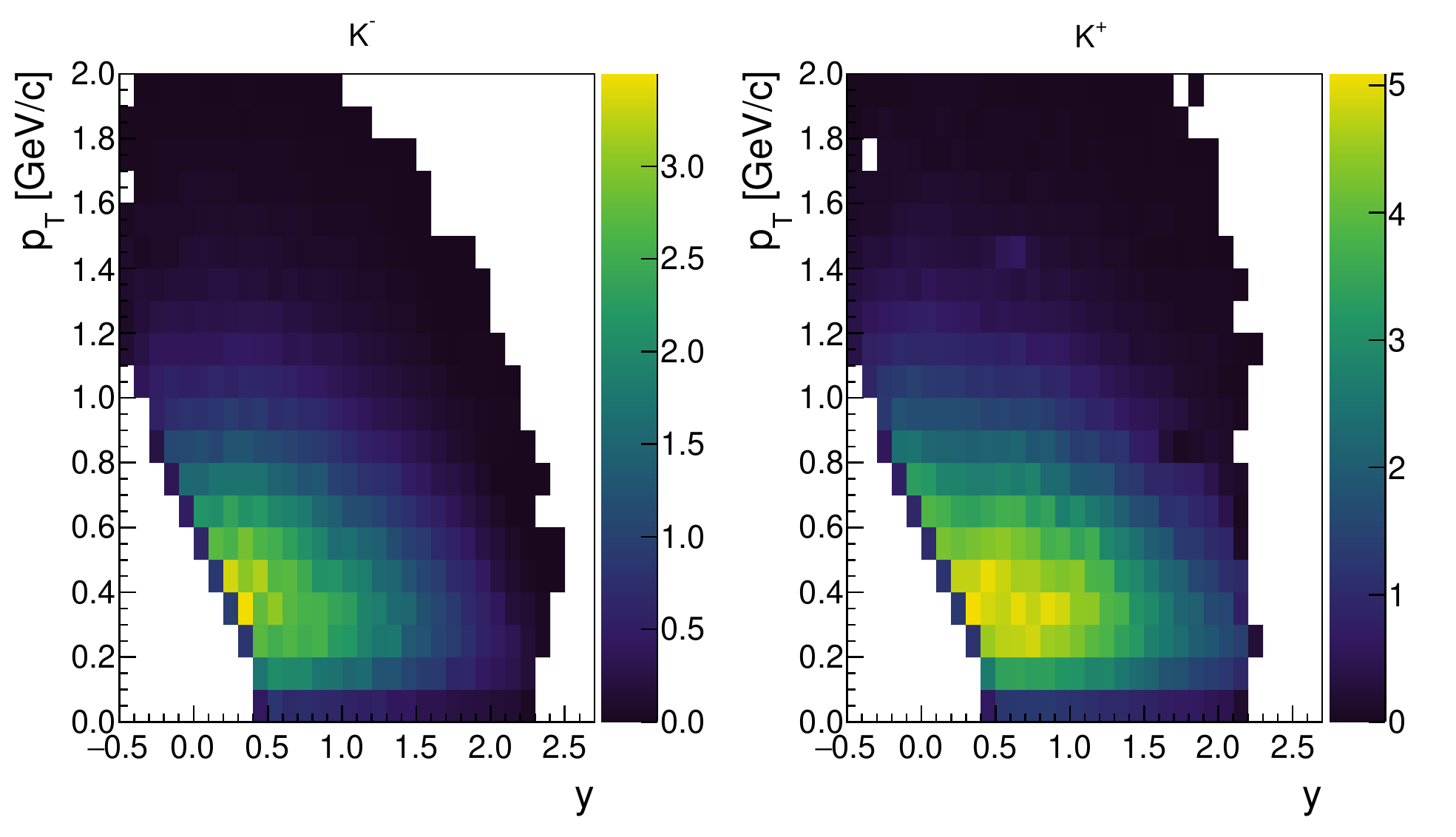}}
\caption{Preliminary double differential yields per event $\frac{d^2n}{dy~dp_T}$ for three beam momenta: 30\textit{A}, 40\textit{A}, 75\textit{A} GeV/\textit{c}.}
\label{fig:ptvy}
\normalsize
\end{figure}

\begin{figure}[h]
\centering{\includegraphics[width=0.25\textwidth]{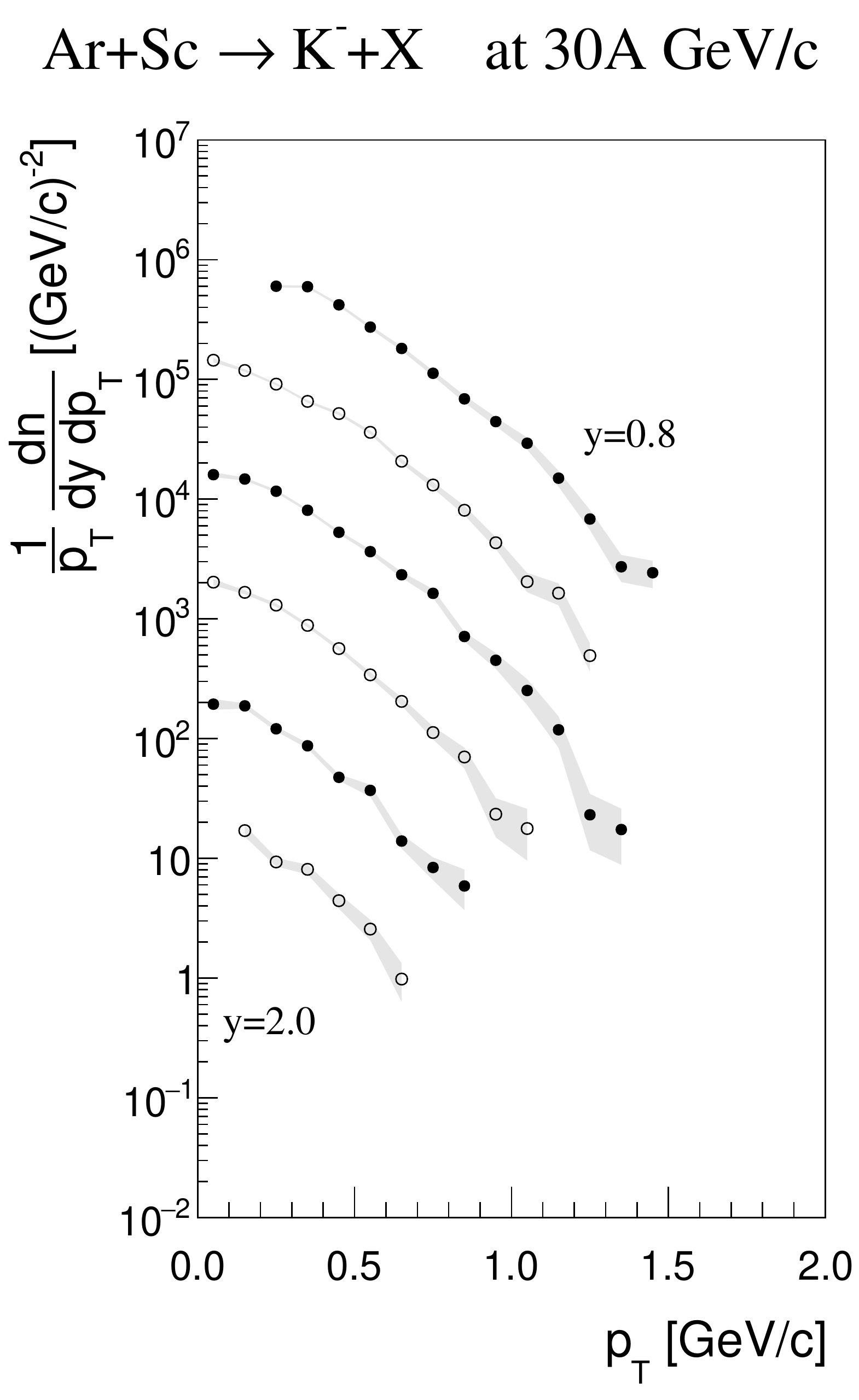}\hspace{0.1cm}
\includegraphics[width=0.25\textwidth]{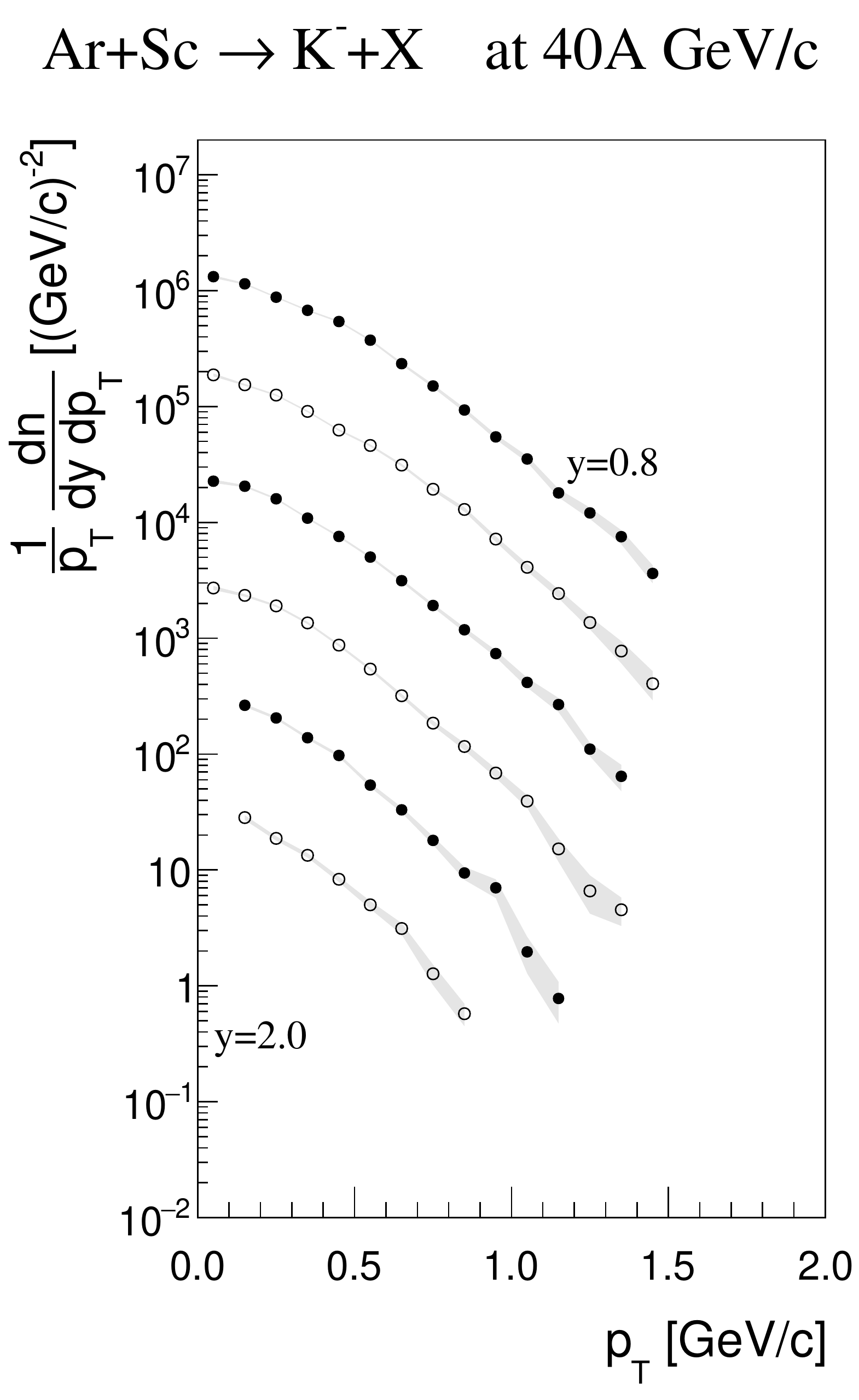}\hspace{0.1cm}
\includegraphics[width=0.25\textwidth]{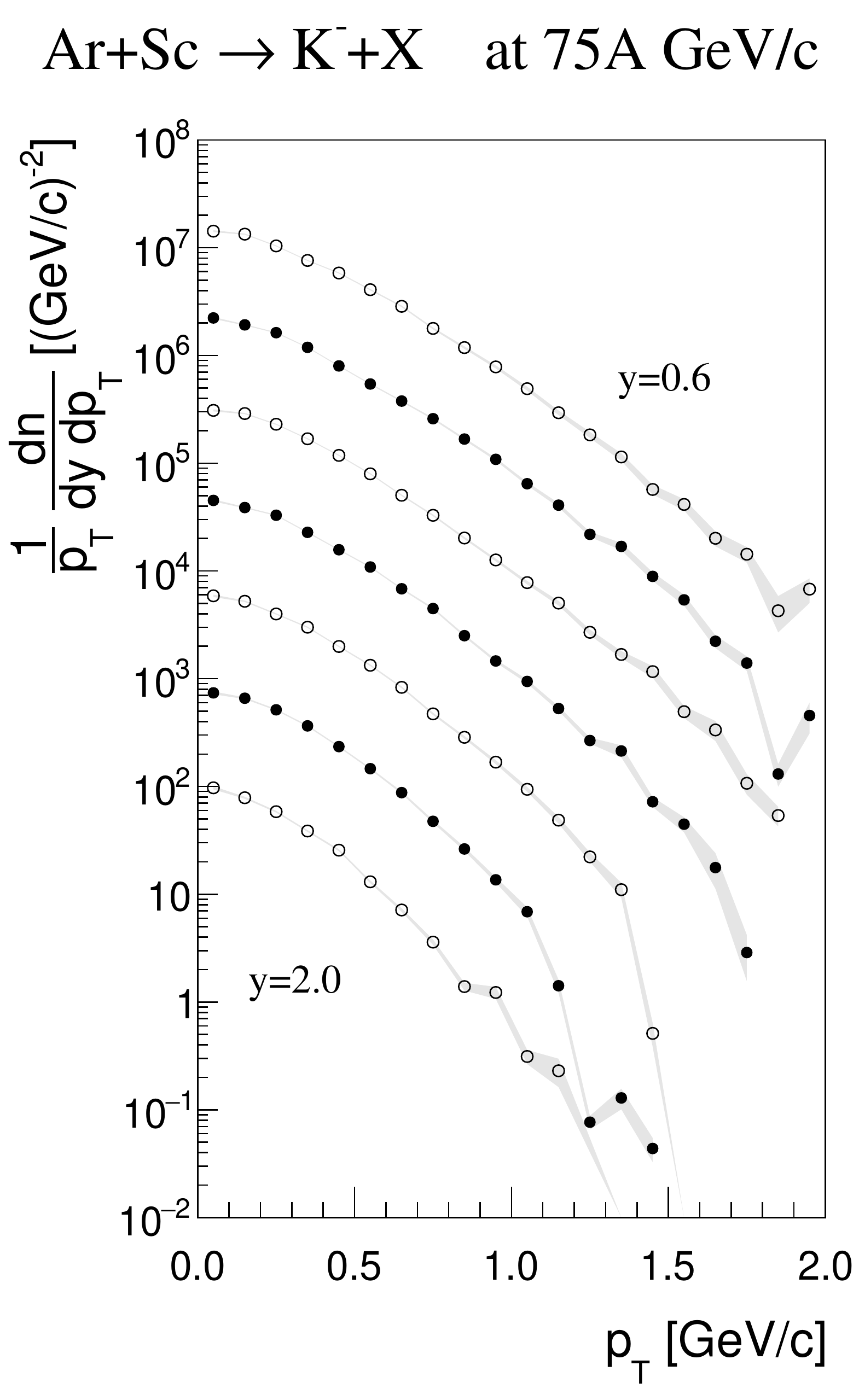}
\includegraphics[width=0.25\textwidth]{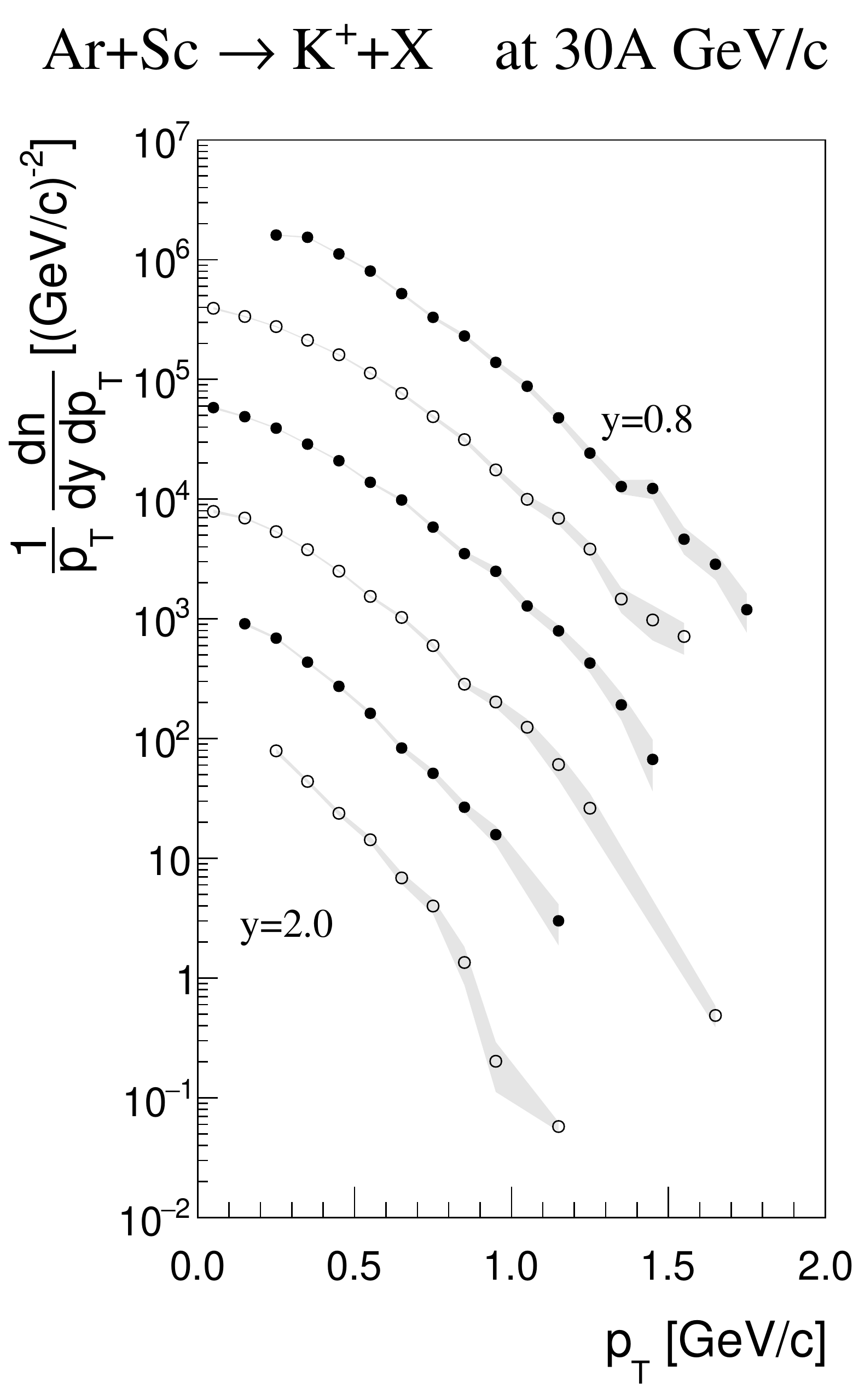}\hspace{0.1cm}
\includegraphics[width=0.25\textwidth]{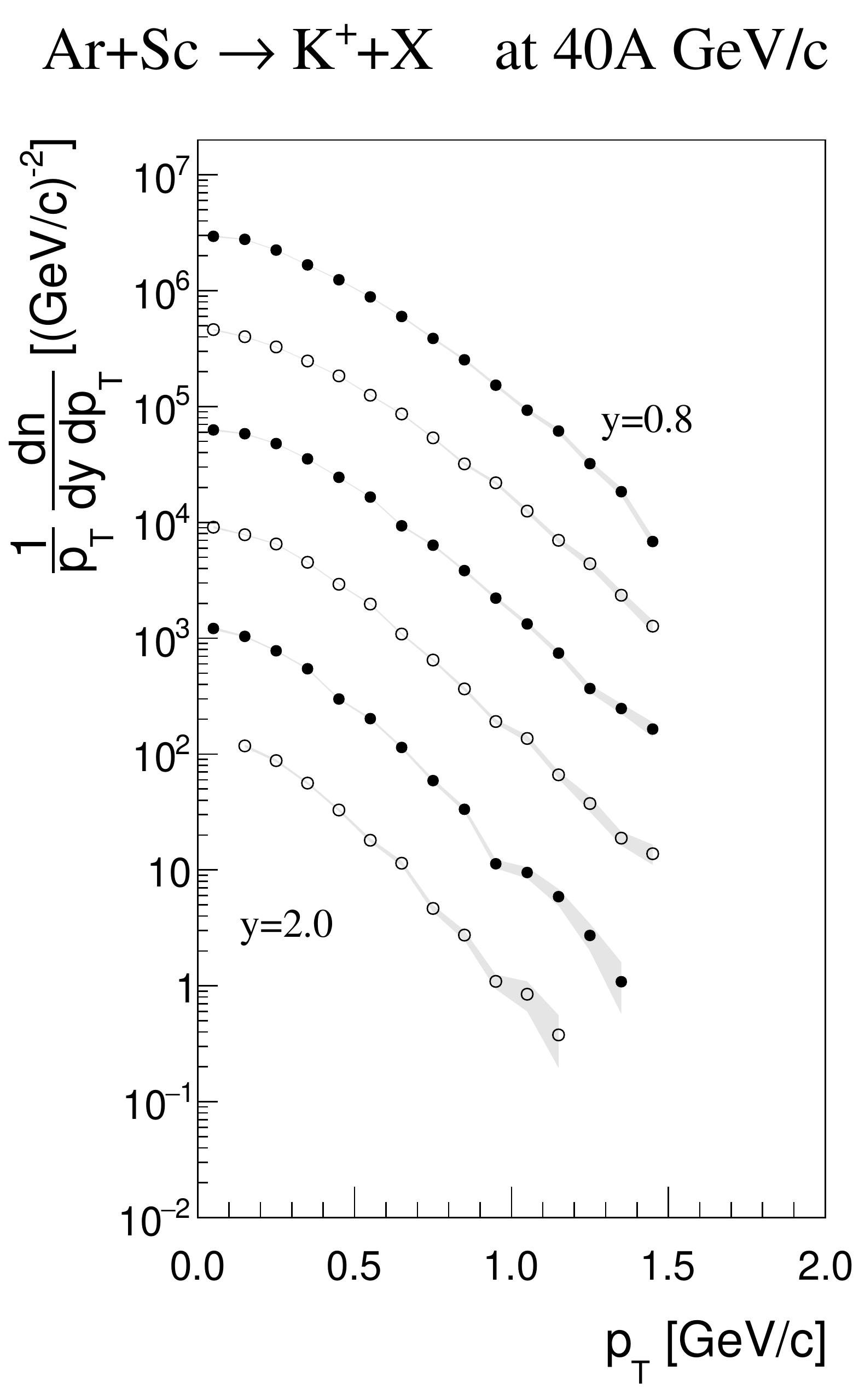}\hspace{0.1cm}
\includegraphics[width=0.25\textwidth]{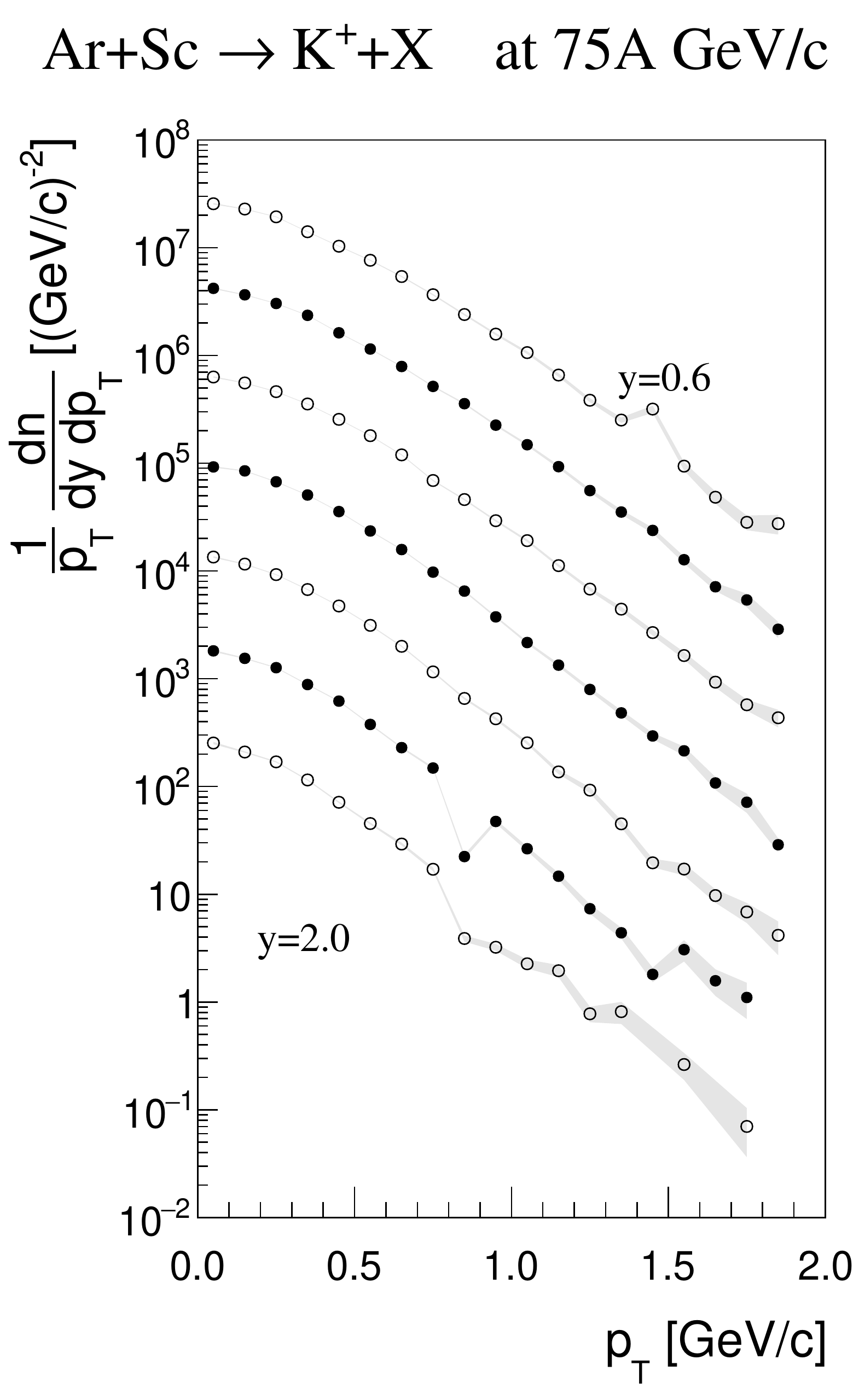}}
\caption{Preliminary transverse momentum spectra $\frac{1}{p_T}\frac{d^2n}{dy~dp_T}$ in slices of rapidity $y$ for three beam momenta: 30\textit{A}, 40\textit{A}, 75\textit{A} GeV/\textit{c}.}
\label{fig:pT}
\end{figure}

Using well measured tracks coming from the primary interaction, double differential yields per event in bins of momentum and transverse momentum for positively and negatively charged kaons were obtained using the $dE/dx$ identification method. The probability method described in the previous section was used to convert the identified particle yields fitted in bins of $(p_{lab}, p_T)$ to spectra in bins $(y, p_T)$ of center-of-mass rapidity and transverse momentum. Fig.~\ref{fig:ptvy} shows the double differential yields per event $\frac{d^2n}{dy~dp_T}$ and Fig.~\ref{fig:pT} displays transverse momentum spectra $\frac{1}{p_T}\frac{d^2n}{dy~dp_T}$ in slices of rapidity~$y$.

\subsection{Extrapolation in $p_T$}
In order to obtain $dn/dy$ yields, the data is extrapolated in $p_T$ to account for unmeasured regions at high values of $p_T$. Exponential dependence in $p_T$ is assumed:
\begin{equation}
\frac{1}{p_T} \frac{d^2n}{dp_T~dy}~~=~~\frac{dn/dy}{T\cdot(m_K+T)}~\cdot~e^{-(m_T-m_K)/T}
\label{eq:pTextr}
\end{equation}
The function is fitted in the acceptance region and its integral beyond the acceptance is added to the measured data. The contribution of the extrapolation is typically of the order of 1\%.

\subsection{Inverse slope parameter $T$}

\begin{figure}[h]
	\includegraphics[width=0.32\textwidth]{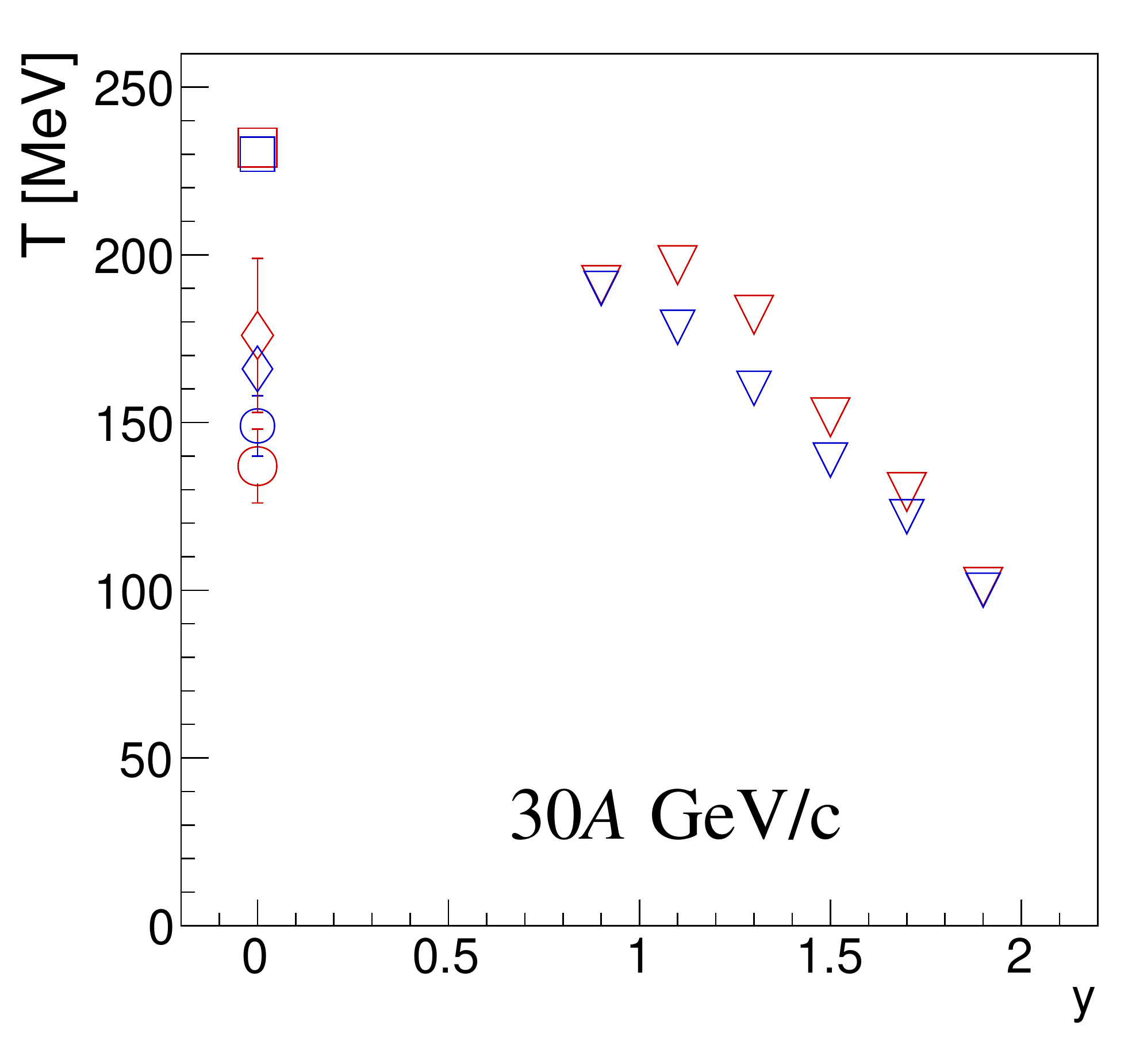} \includegraphics[width=0.32\textwidth]{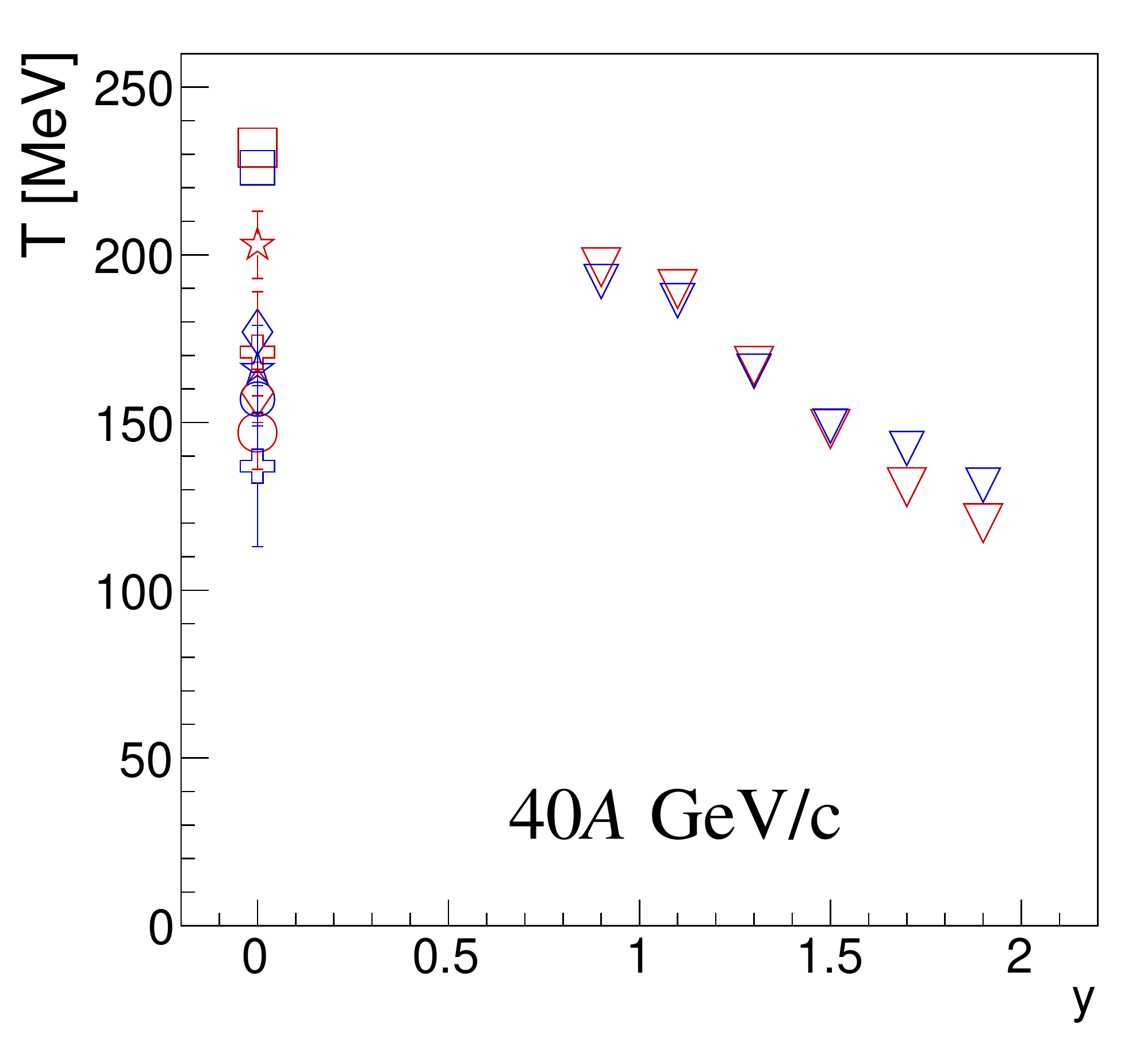}
	\includegraphics[width=0.32\textwidth]{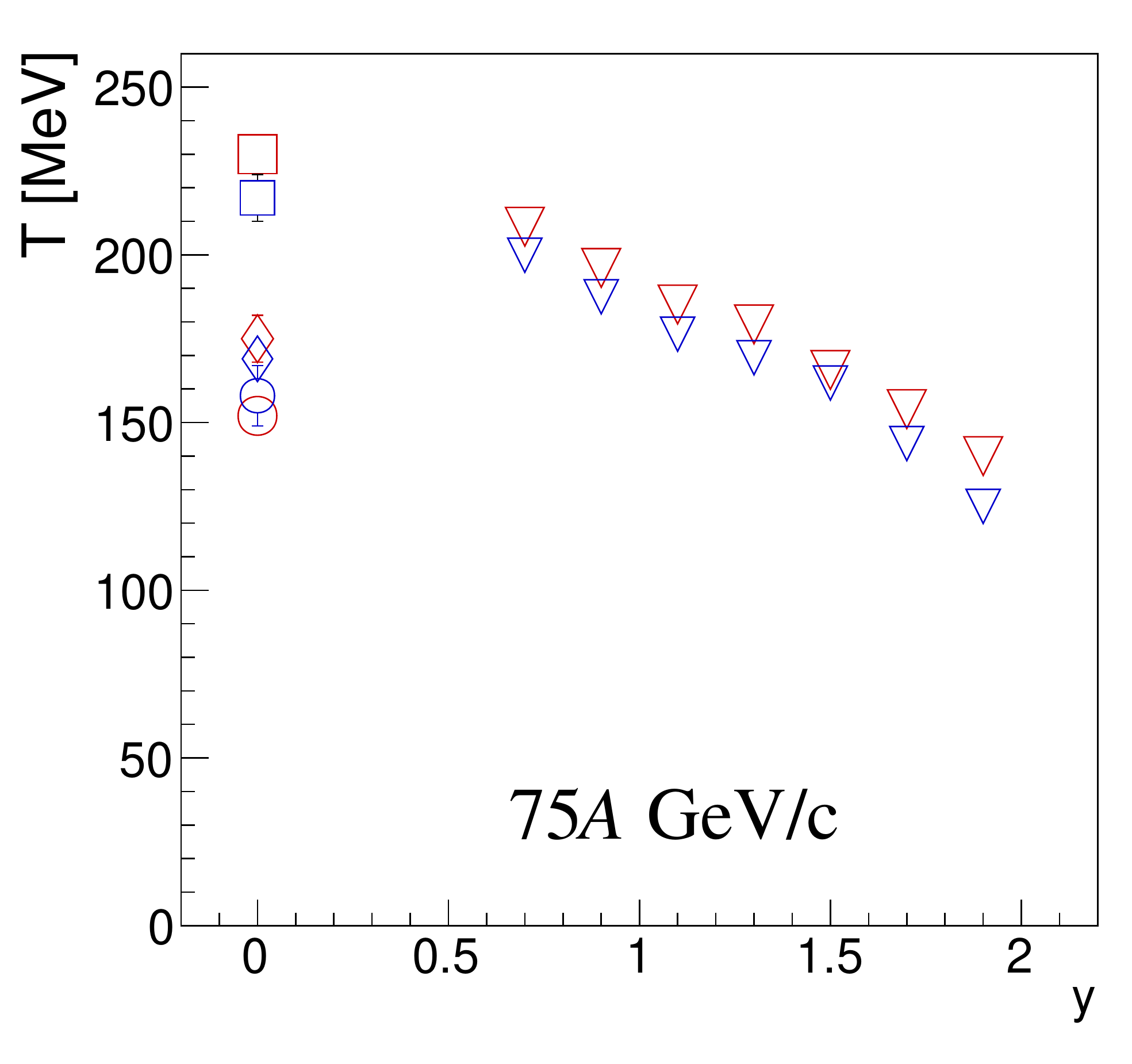}\\
	\tikzmark{mark}
	\begin{tikzpicture}[overlay, remember picture]
	\node (one) at ($(mark.south)+(1.2,0.0)$) {};
	\node (two) at ($(one.south)+(6.7,-0.8)$) {};
	\node[at=(two)] {
		\includegraphics[width=0.46\paperwidth]{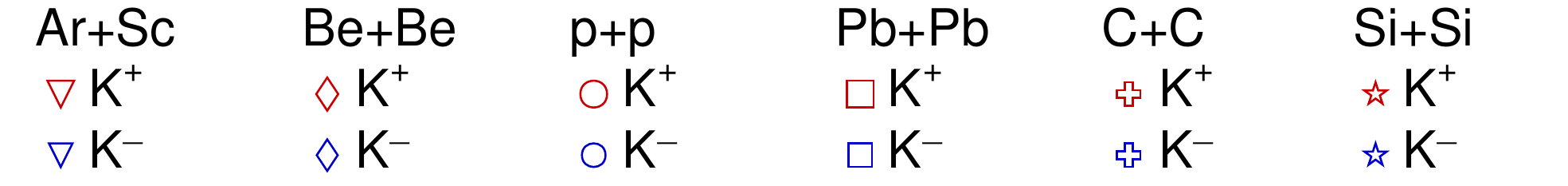}
	};
	\small
	\draw [decoration={brace,amplitude=5pt},decorate,ultra thick,gray]
	($(one.south)+(2,0.0)$) -- ($(one.south)+(6.1,0.0)$) node [black,midway,yshift=13pt,align=left] {NA61/SHINE};
	\draw [decoration={brace,amplitude=5pt},decorate,ultra thick,gray]
	($(one.south)+(6.9,0.0)$) -- ($(one.south)+(11.1,0.0)$) node [black,midway,yshift=13pt,align=left] {NA49};
	\draw [decoration={brace,amplitude=5pt},decorate,thick,gray]
	($(one.south)+(4.6,-1.4)$) -- ($(one.south)+(2.0,-1.4)$) node [black,midway,yshift=-13pt,align=left] {Preliminary};
	\end{tikzpicture}
	\vspace{2cm}
	\caption{Inverse slope parameter $T$ of transverse momentum spectra. The reversed triangles denote preliminary \s{Ar+Sc} results. Other symbols show for comparison results for $T$ for other systems from NA61/SHINE and NA49. The extrapolations to mid-rapidity suggest similar values for \s{Ar+Sc} and for \s{Pb+Pb}. }
	\label{fig:slope}
\end{figure}
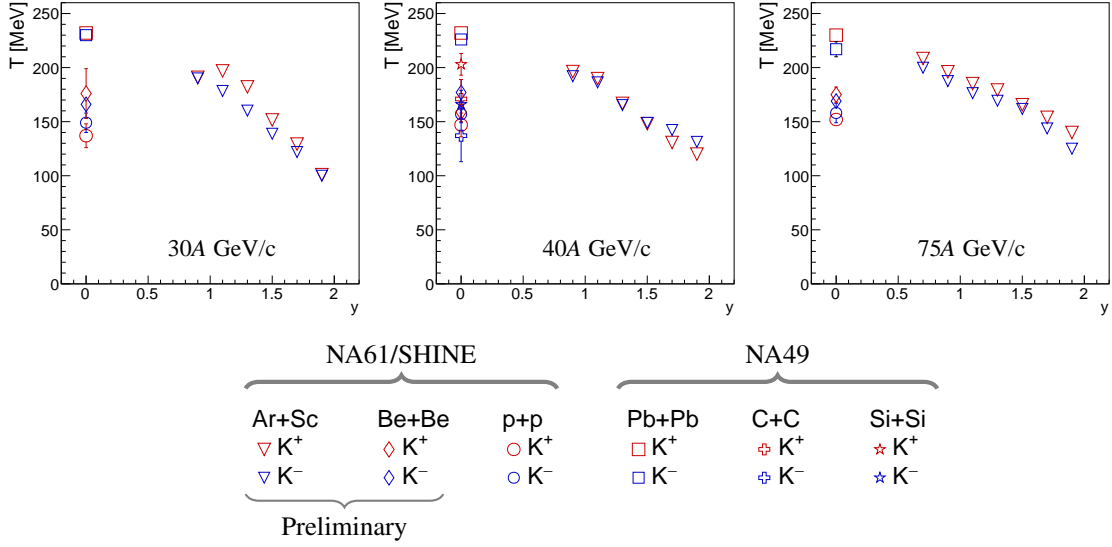

The fit of transverse momentum spectra with Eq.~\ref{eq:pTextr} determines the inverse slope parameter $T$. The results obtained for the three analyzed collision energies are shown in Fig.~\ref{fig:slope}. An extrapolation of the \s{Ar+Sc} data points to $y\approx0$ approaches the values obtained for \s{Pb+Pb} collisions. Smaller systems (\s{p+p}, \s{Be+Be}, \s{C+C} and \s{Si+Si}) show significantly lower values of~$T$.

\section{Rapidity Spectra}
The double differential spectra $\frac{d^2n}{dy~dp_T}$ described in the previous section are integrated with respect to transverse momentum $p_T$ to calculate the one-dimensional distribution $\frac{dn}{dy}$ of rapidity. The data points are obtained in the acceptance region of the $dE/dx$ particle identification method (forward-rapidity): $y\in[0.8;2.0]$ for 30\textit{A} and 40\textit{A} GeV/\textit{c} beam momenta and a slightly wider range $y\in[0.6;2.0]$ for 75\textit{A} GeV/\textit{c}. 

In order to obtain the $4\pi$ full phase space mean kaon multiplicity, the $\frac{dn}{dy}$ spectra are interpolated in the mid-rapidity region. 

\begin{figure}[h]

\centering{
	\hspace{0.05\textwidth} 30\textit{A} GeV/\textit{c} \hspace{0.18\textwidth} 40\textit{A} GeV/\textit{c} \hspace{0.2\textwidth} 75\textit{A} GeV/\textit{c} \\
	\includegraphics[width=0.32\textwidth]{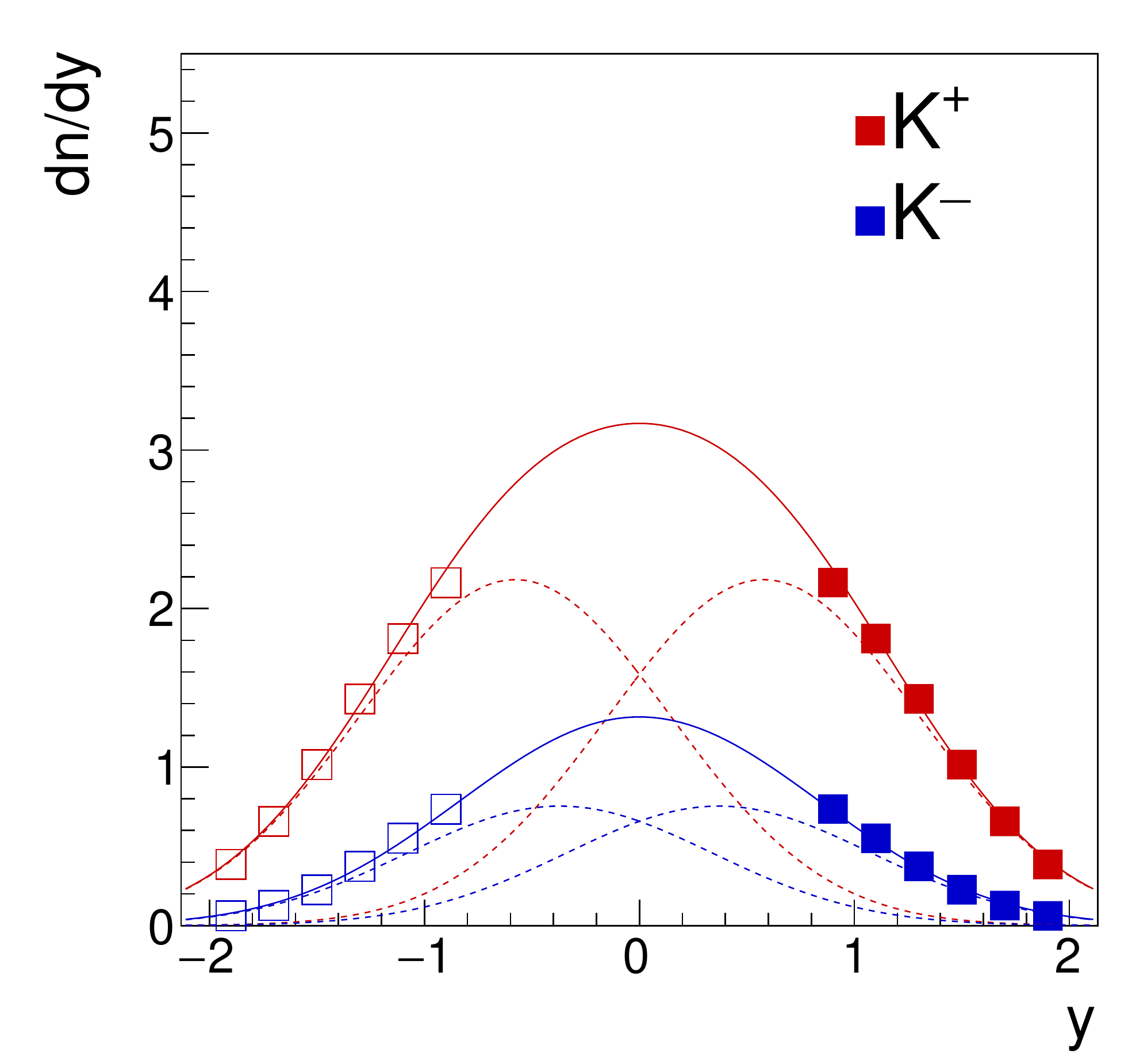}
	\includegraphics[width=0.32\textwidth]{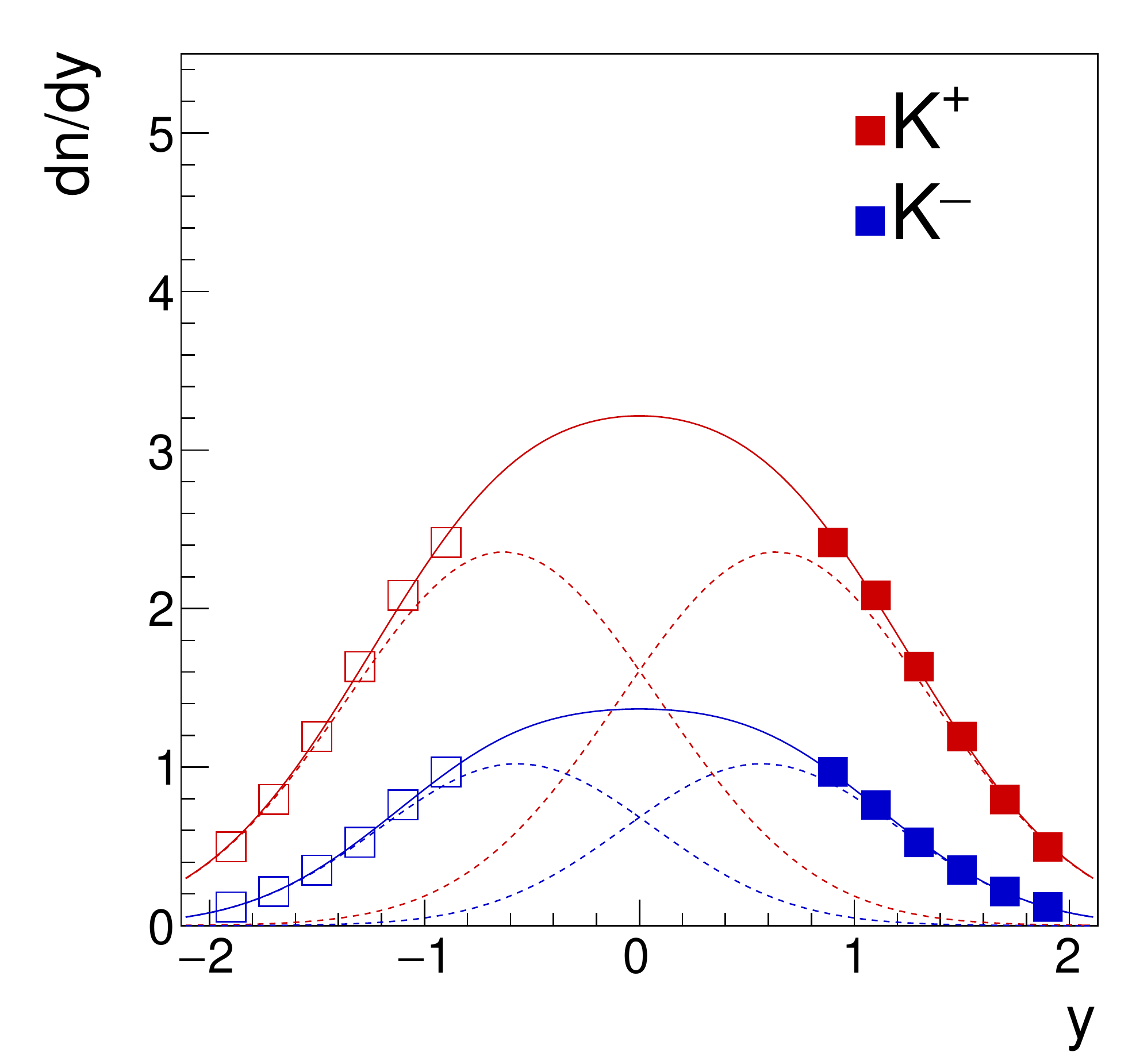}
	\includegraphics[width=0.32\textwidth]{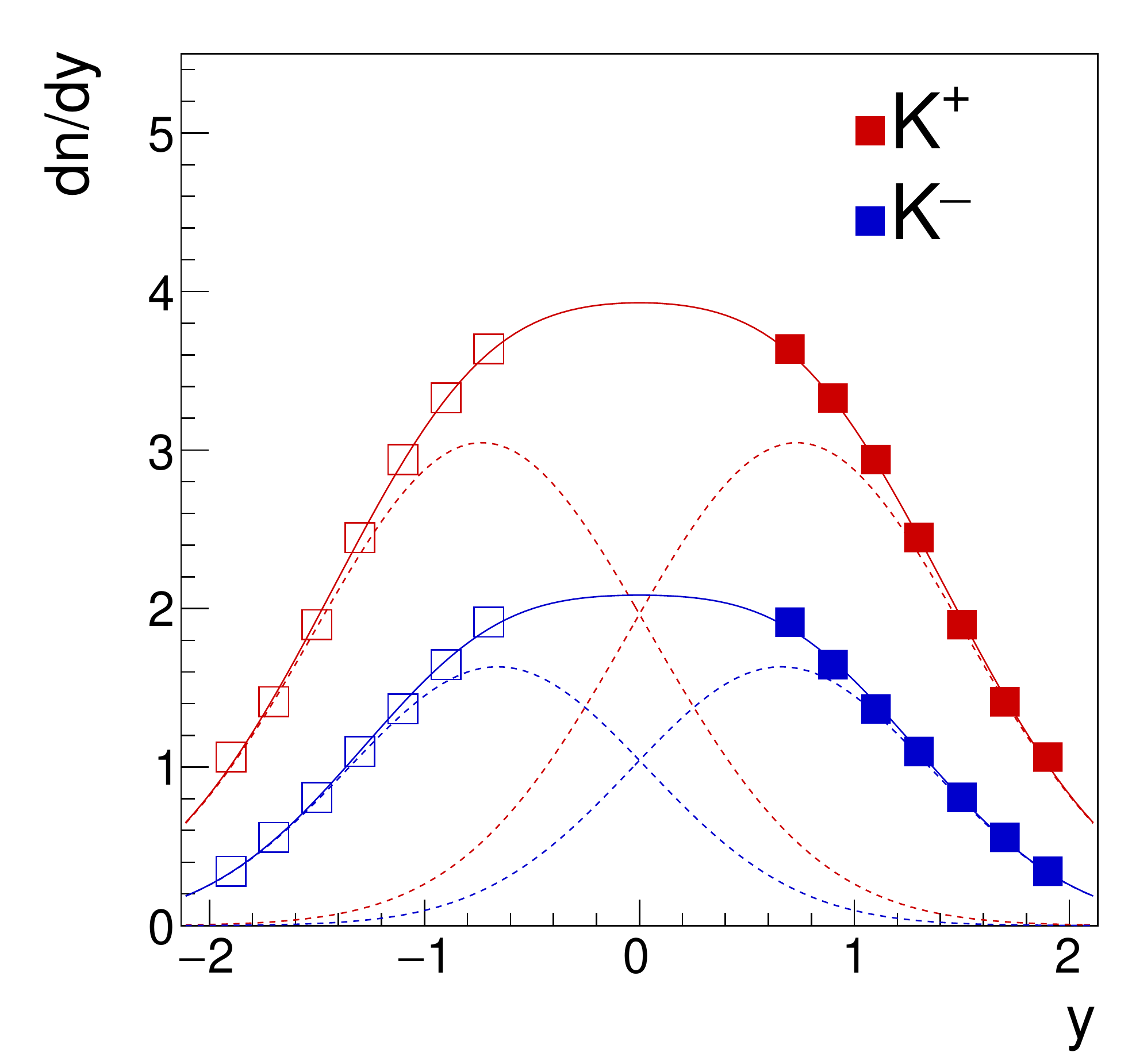}}
 	\caption{Rapidity distributions $\frac{dn}{dy}$ (preliminary). Full squares show measurements and empty data points mark their reflection with respect to mid-rapidity. Curves show a fit (see text) with shape parameters $\sigma$ and $y_0$ fixed to values obtained from \s{Pb+Pb} collisions. }

\label{fig:ydist}

\end{figure}

For this purpose the measured points of the rapidity distributions are reflected with respect to $y=0$ and a fit is performed with two symmetrically displaced Gaussian functions (see Fig.~\ref{fig:ydist}):

\begin{equation}
f_{fit}(y) = \frac{A}{\sigma \sqrt{2\pi}} \exp \left( -\frac{(y-y_0)^2}{2\sigma^2} \right) + \frac{A}{\sigma \sqrt{2\pi}} \exp \left( -\frac{(y+y_0)^2}{2\sigma^2} \right)
\label{eq:yfit}
\end{equation}

The limited range of the measurements does not allow a fit with unconstrained parameters ($A,y_0,\sigma$). As a significant similarity of rapidity distributions from \s{Ar+Sc} and \s{Pb+Pb} collisoins was observed, the shape parameters $y_0$ and $\sigma$ were fixed to the values obtained for kaon spectra measurements in \s{Pb+Pb} reactions~\cite{NA49PbPb,NA49PbPb2}.

To account for minor differences of the compared beam momenta (\s{Ar+Sc} at 75\textit{A} and \s{Pb+Pb} at 80\textit{A} GeV/\textit{c}) the following scaling of parameters was applied:

\begin{equation}
\sigma = \frac{y^{(beam)}_{75}}{y^{(beam)}_{80}} \cdot \sigma_{(PbPb@80)},~~~y_0 = \frac{y^{(beam)}_{75}}{y^{(beam)}_{80}} \cdot y_{0~{(PbPb@80)}}
\end{equation}

where $y^{(beam)}_{p}$ is the beam rapidity for beam momentum $p$. Thus the amplitude $A$ is the only free parameter and is fitted to both the measured and reflected parts of the spectra. It is remarkable how well the shape parameters from \s{Pb+Pb} collisions fit the \s{Ar+Sc} data (see Fig.~\ref{fig:ydist}).

\section{Mean Kaon Multiplicities}

In order to obtain the mean kaon multiplicities $\langle K^+\rangle$ and $\langle K^-\rangle$, the rapidity spectra $\frac{dn}{dy}$ are integrated by taking the sum of measurements in the acceptance region (forward-rapidity) and the integral of the fitted function outside the acceptance (mid-rapidity and $y>2.0$). Such a calculation is clearly laden with significant uncertainty:

\begin{itemize}\setlength\itemsep{-0.1cm}

	\item statistical error (calculated precisely, originating from data and fit procedure)

	\item uncertainty due to the choice of fit function parameters.

\end{itemize}

The latter was estimated by varying the shape parameters $y_0$ and $\sigma$. Obtained uncertainty estimates had typical values of $\lesssim10\%$ and are shown by orange-shaded bars in Figs.~\ref{fig:horn}-\ref{fig:kpn}. A better estimate of systematic uncertainties will be possible in future when complementary measurements of \textit{tof} will allow to obtain data points in the mid-rapidity region.

\subsection{System size dependence of kaon production}

\begin{figure}[h]

\centering{\includegraphics[width=0.6\textwidth]{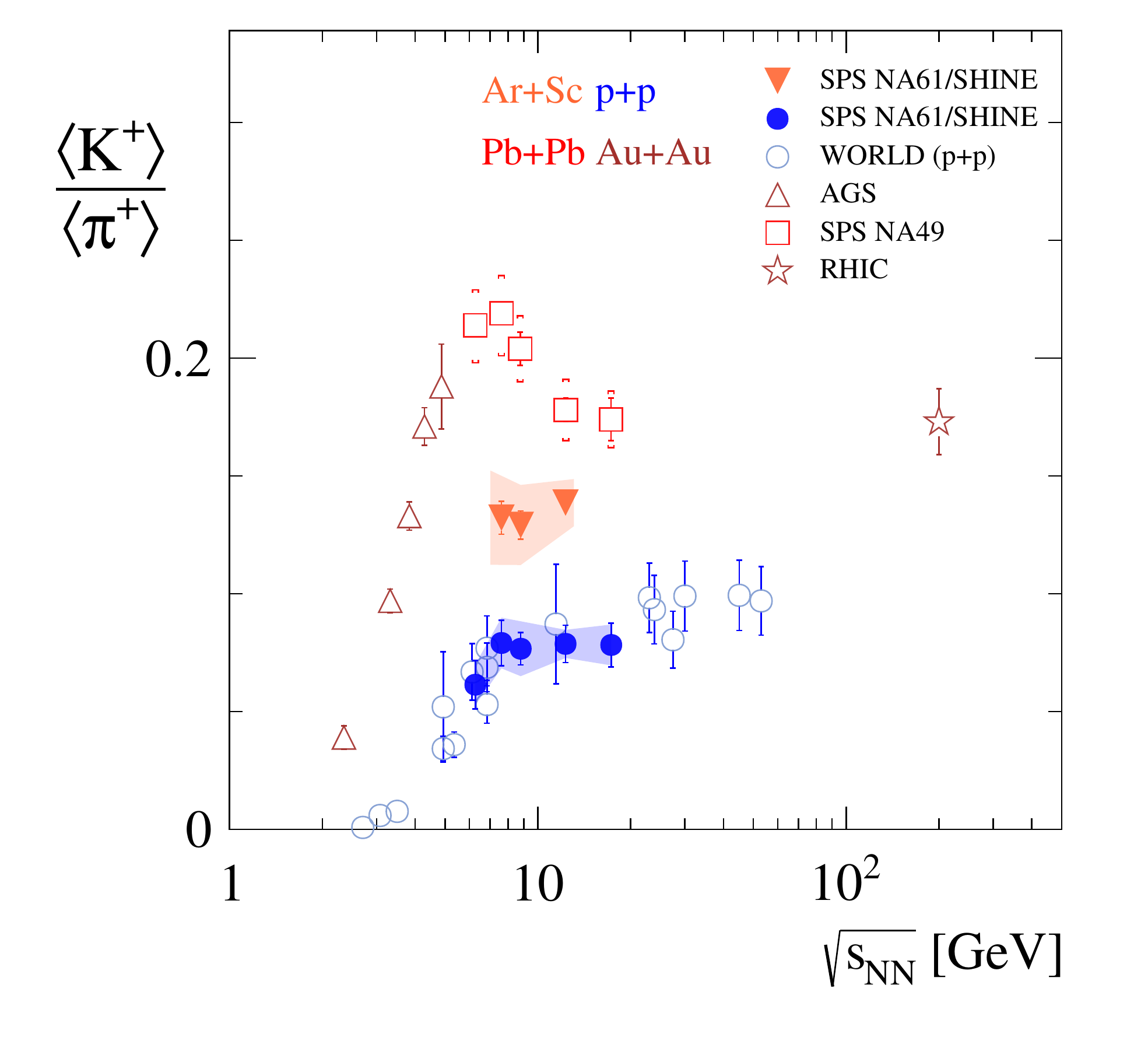}}

\caption{Energy dependence of $\left<K^{+}\right>/\left<\pi^{+}\right>$ -- {"the horn" plot}.}

\label{fig:horn}

\end{figure}

Particle production changes rapidly with collision energy in the vicinity of the onset of deconfinement. One observes a clear, qualitative difference between results obtained in \s{p+p} collisions and collisions of heavy nuclei (\s{Pb+Pb} or \s{Au+Au}). This difference is especially pronounced in the production of strange hadrons as exemplified by the \textit{horn} plot shown in Fig.~\ref{fig:horn}.


\begin{figure}[h]
\includegraphics[width=0.32\textwidth]{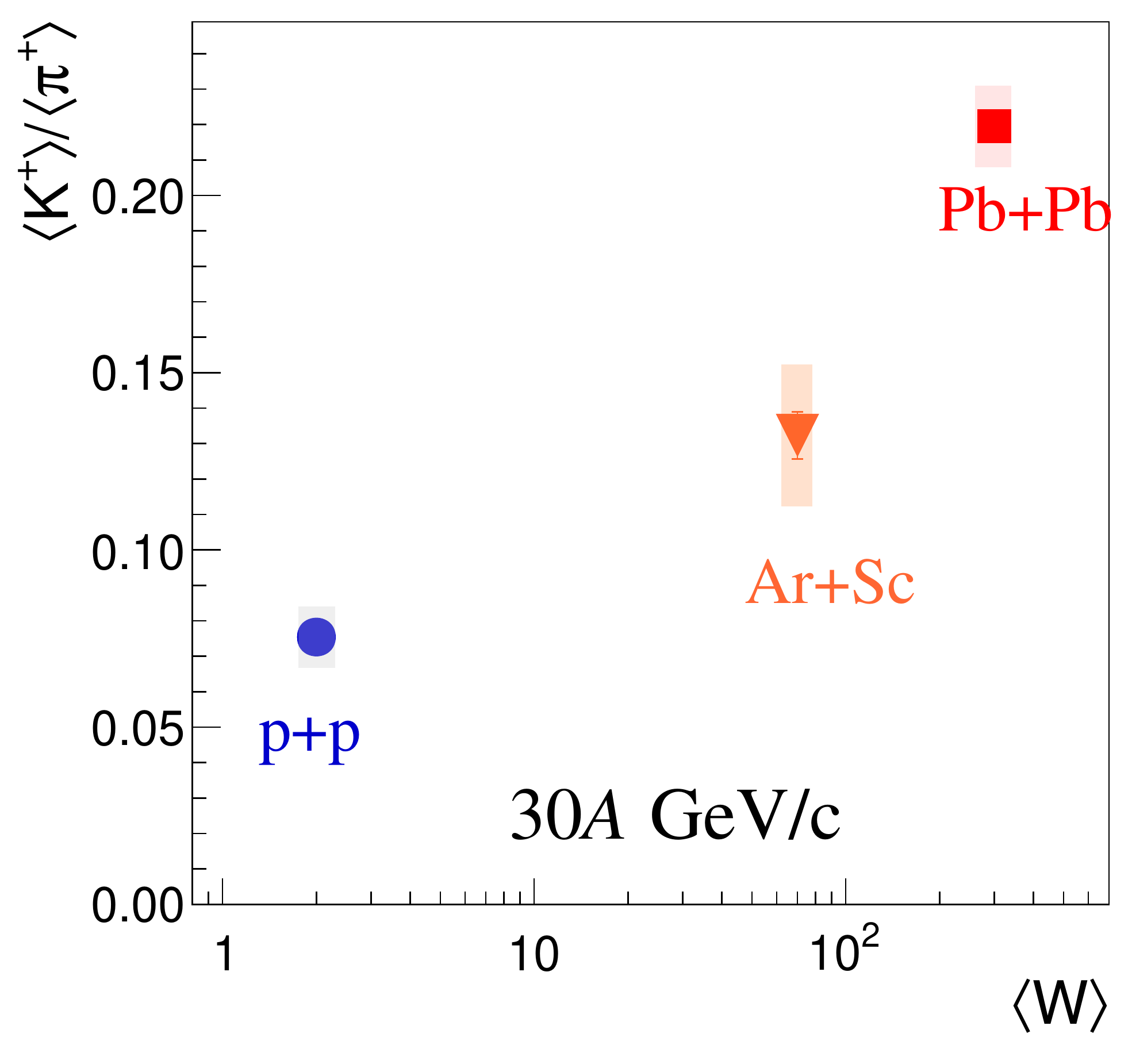}
\includegraphics[width=0.32\textwidth]{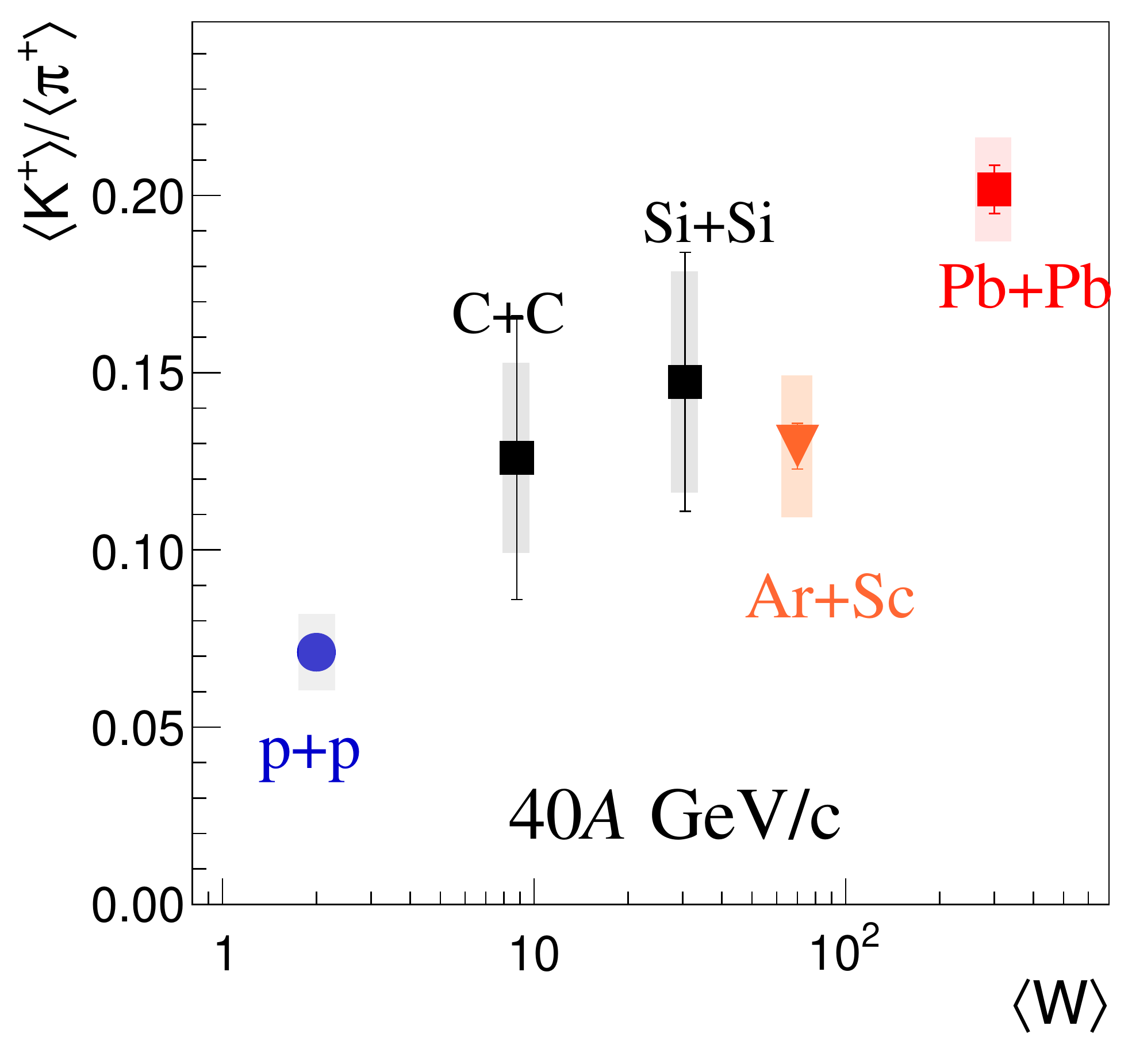}
\includegraphics[width=0.32\textwidth]{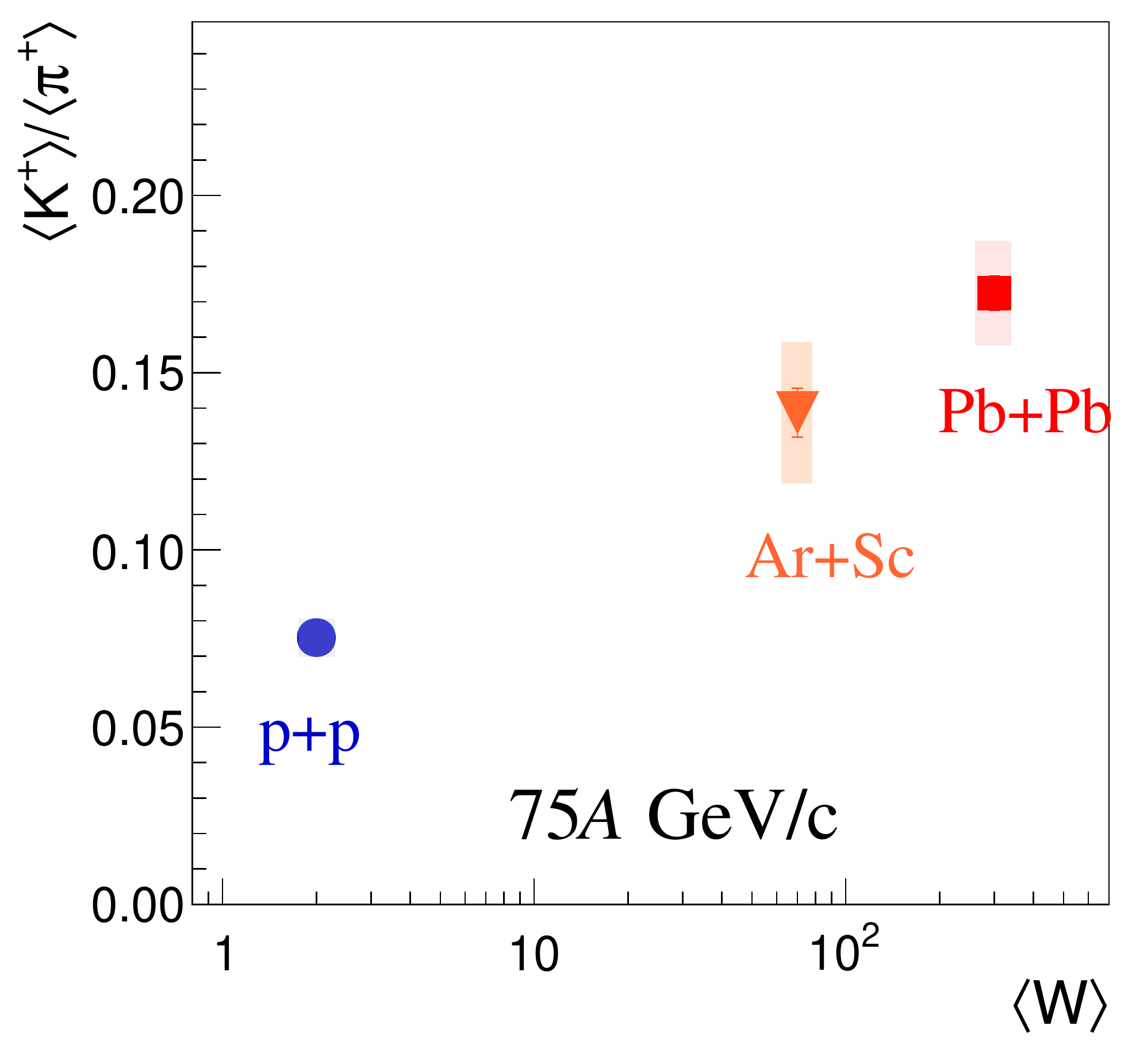}
\caption{System size dependence of the $\left<K^{+}\right>/\left<\pi^{+}\right>$ ratio for central \s{Ar+Sc} collisions (preliminary NA61/SHINE measurements), \s{p+p} interactions (NA61/SHINE), as well as central \s{Pb+Pb} and semi-central \s{C+C} and \s{Si+Si} collisions (NA49 Ref.~\cite{NA49CCSiSi}).}

\label{fig:kpi}

\end{figure}

\begin{figure}[h]
\includegraphics[width=0.32\textwidth]{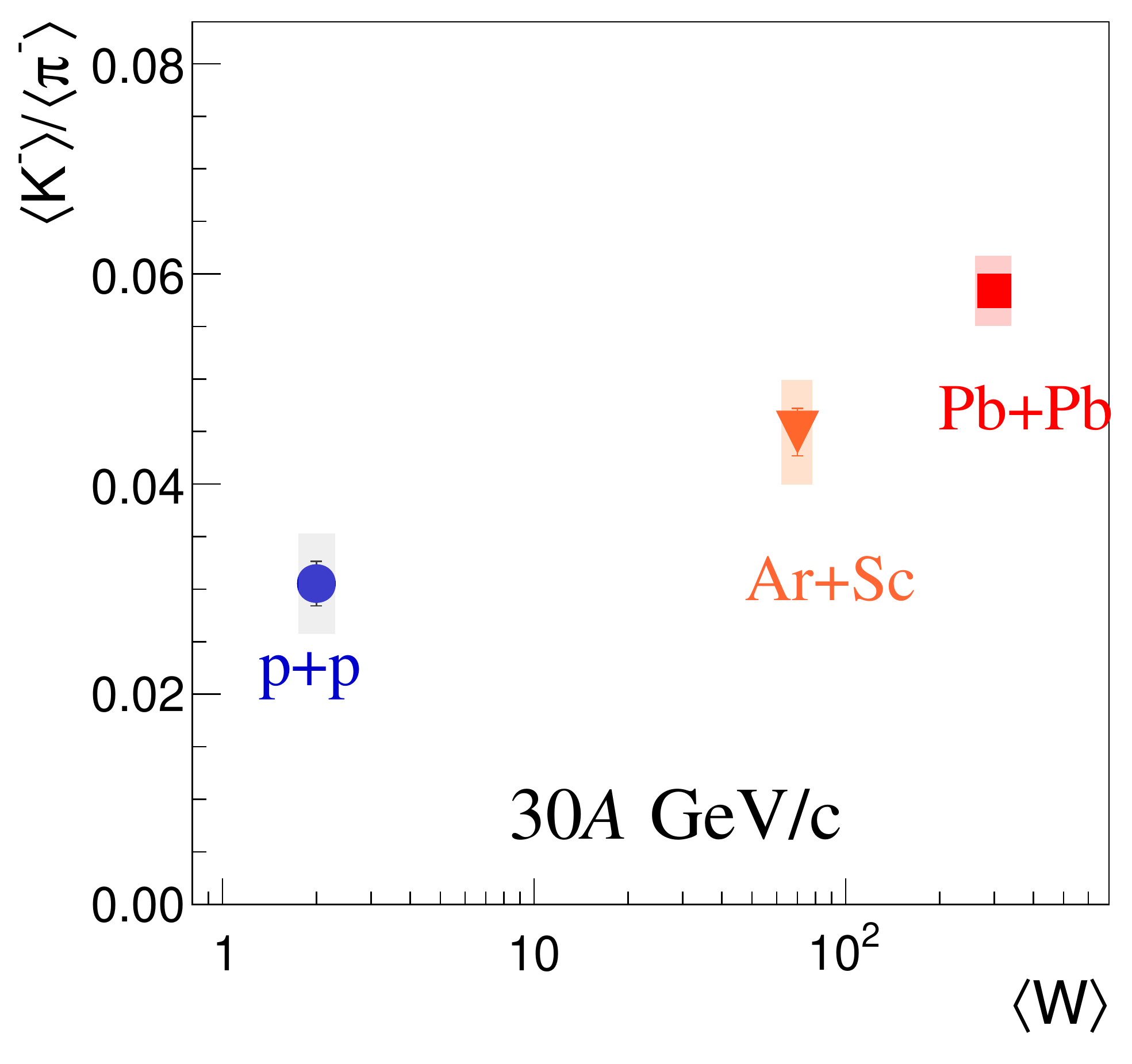}
\includegraphics[width=0.32\textwidth]{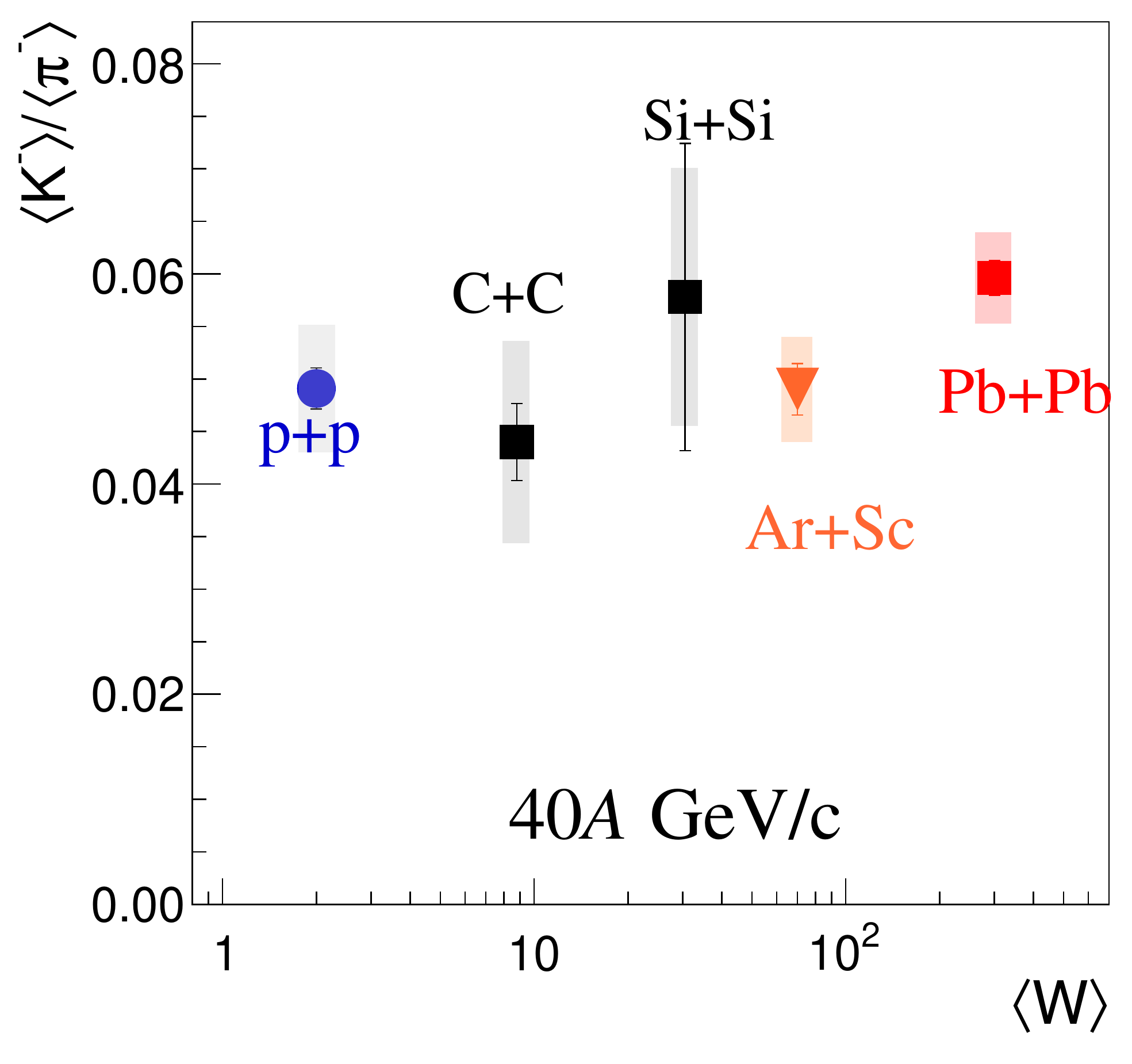}
\includegraphics[width=0.32\textwidth]{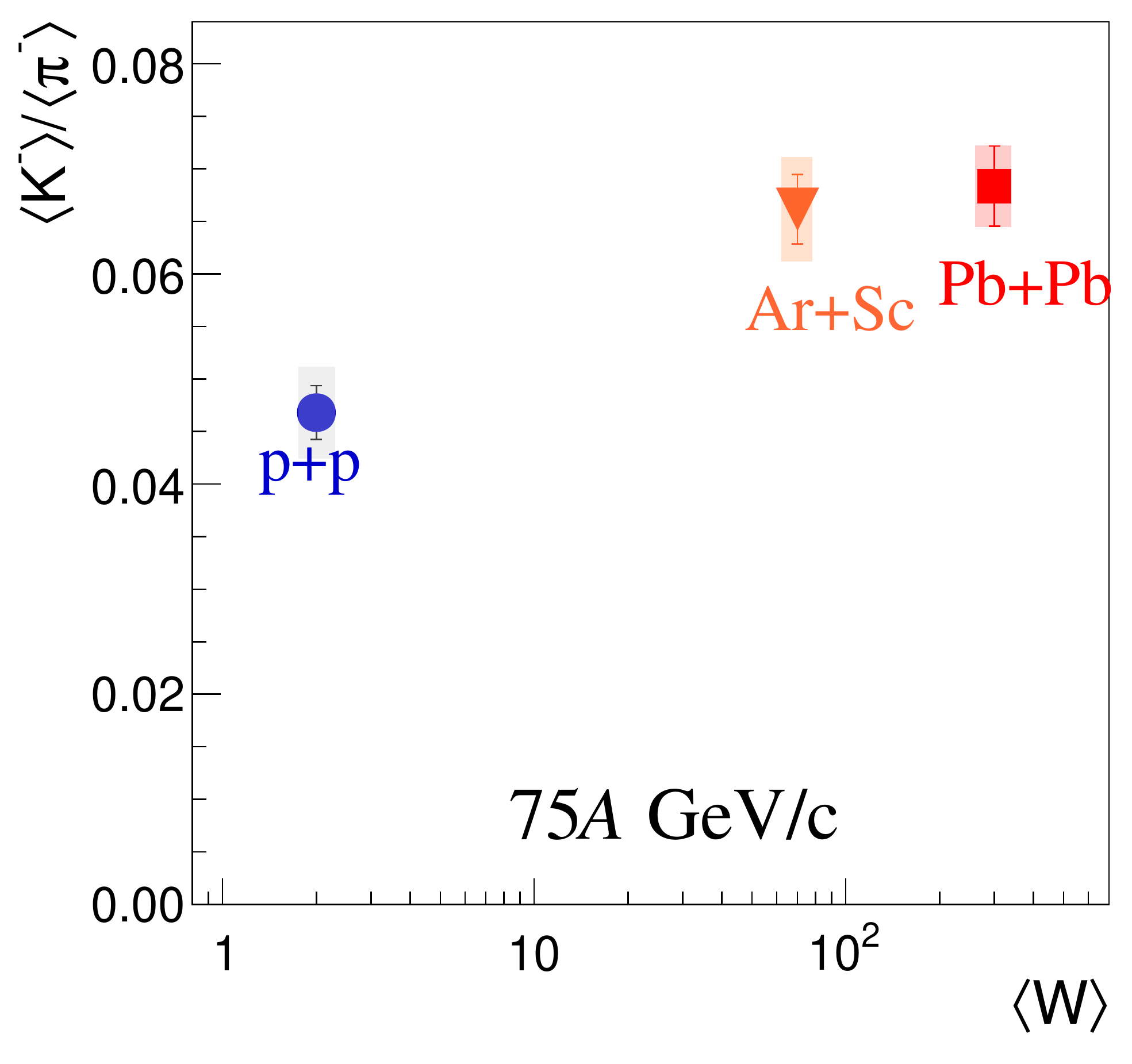}
\caption{System size dependence of the $\left<K^{-}\right>/\left<\pi^{-}\right>$ ratio for central \s{Ar+Sc} collisions (preliminary NA61/SHINE measurements), \s{p+p} interactions (NA61/SHINE), as well as central \s{Pb+Pb} and semi-central \s{C+C} and \s{Si+Si} collisions (NA49 Ref.~\cite{NA49CCSiSi}).}

\label{fig:nkpi}

\end{figure}

\begin{figure}[h]
\includegraphics[width=0.32\textwidth]{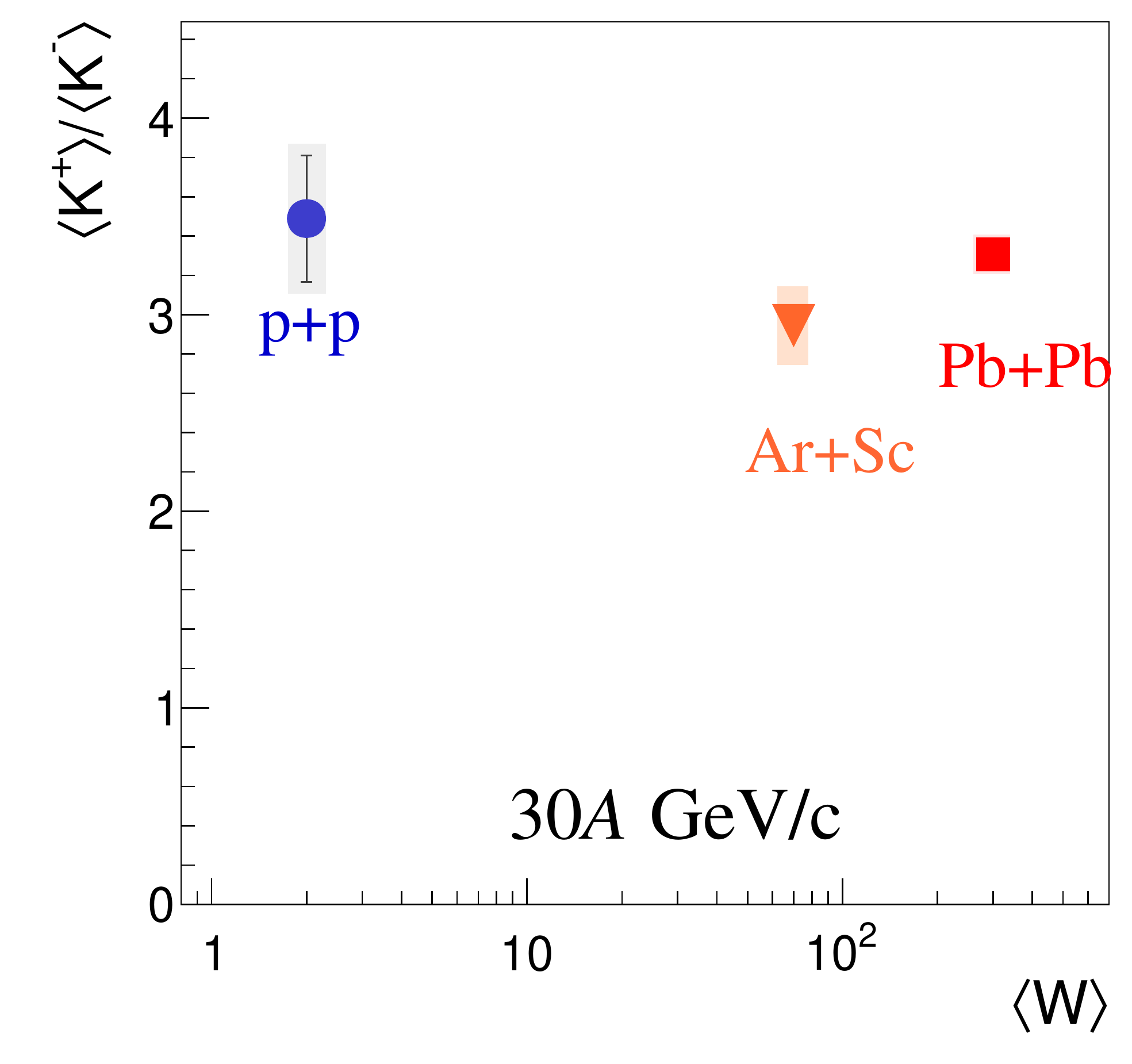}
\includegraphics[width=0.32\textwidth]{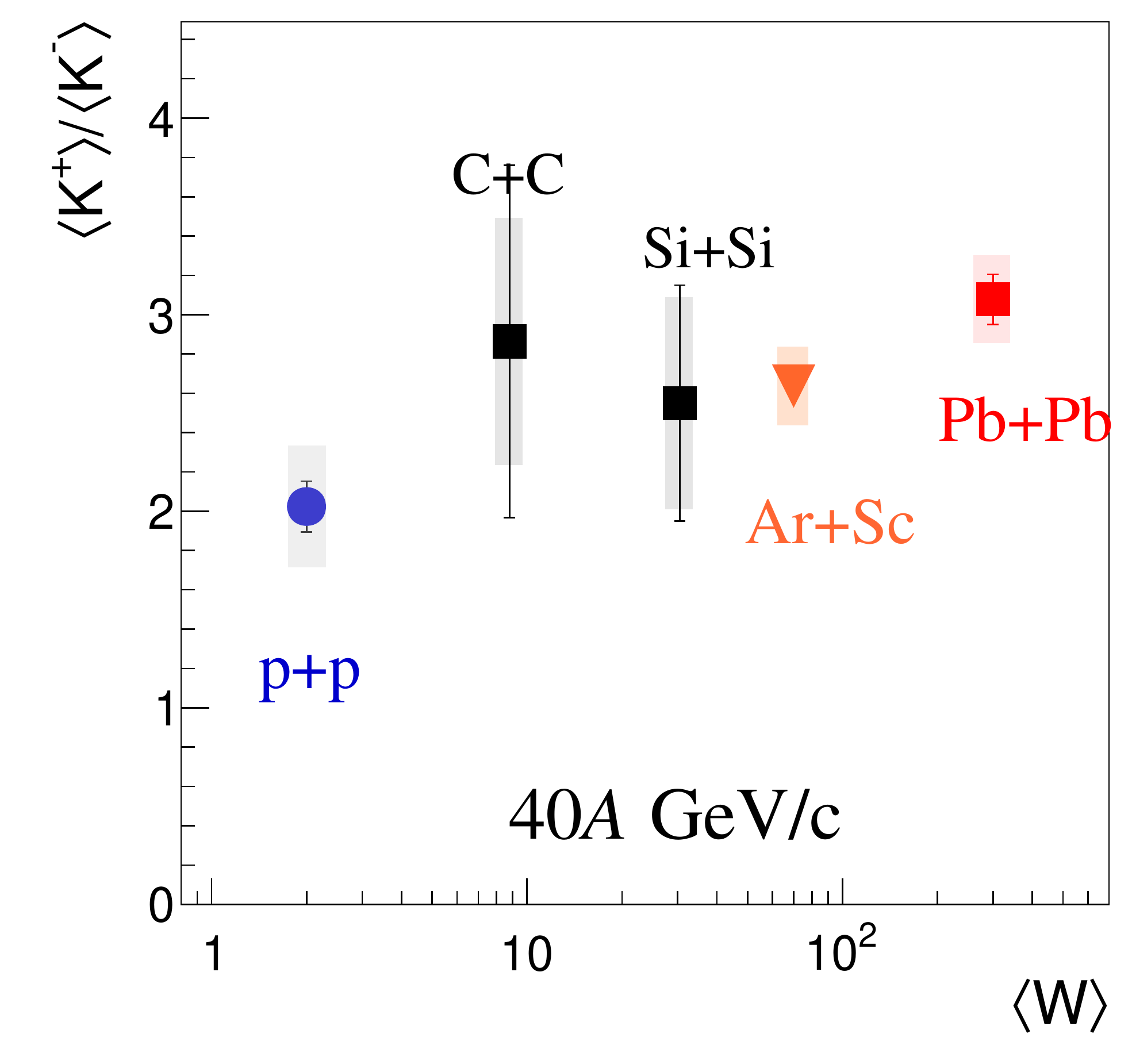}
\includegraphics[width=0.32\textwidth]{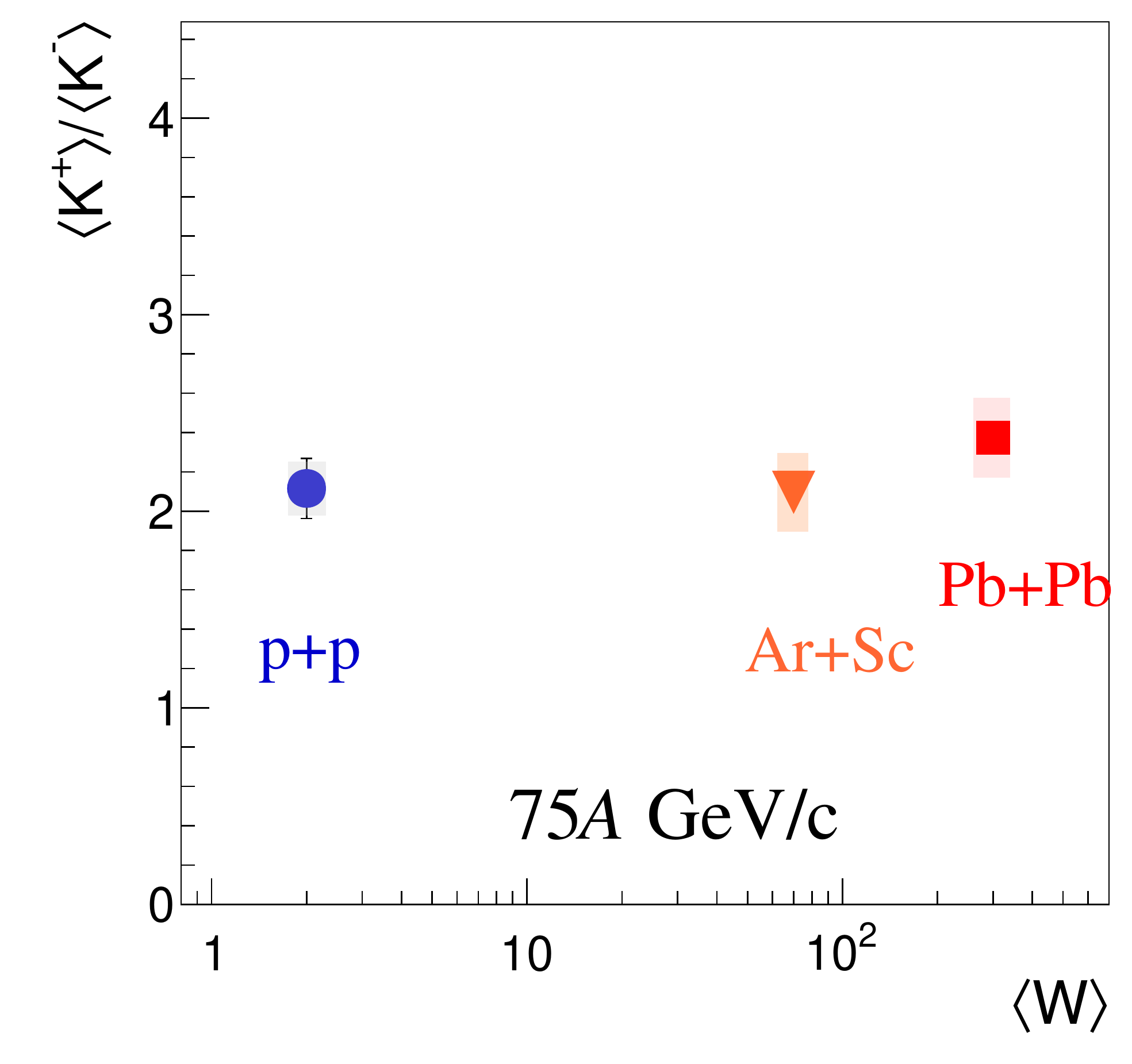}
\caption{{System size dependence of the $\left<K^{+}\right>/\left<K^{-}\right>$} ratio for central \s{Ar+Sc} collisions (preliminary NA61/SHINE measurements), \s{p+p} interactions (NA61/SHINE), as well as central \s{Pb+Pb} and semi-central \s{C+C} and \s{Si+Si} collisions (NA49 Ref.~\cite{NA49CCSiSi}).}

\label{fig:kpn}

\end{figure}

The ratios of mean multiplicites $\left<K^{+}\right>/\left<\pi^{+}\right>,~\left<K^{-}\right>/\left<\pi^{-}\right>$ and $\left<K^{+}\right>/\left<K^{-}\right>$ measured in central \s{Ar+Sc} collisions at beam momenta of 30\textit{A}, 40\textit{A} and 75\textit{A} GeV/\textit{c} are compared in Figs.~\ref{fig:kpi}-\ref{fig:kpn} with results from other systems:

\begin{itemize}\setlength\itemsep{-0.1cm}

	\item \s{p+p} interactions (30, 40, 80 GeV/\textit{c})~\cite{SZpp}

	\item central \s{Pb+Pb} collisions (30\textit{A}, 40\textit{A}, 80\textit{A} GeV/\textit{c})~\cite{NA49PbPb,NA49PbPb2}

	\item semi-central \s{C+C}, \s{Si+Si} collisions (40\textit{A} GeV/\textit{c})~\cite{NA49CCSiSi}

\end{itemize}

In the calculation of the $\left<K^{+}\right>/\left<\pi^{+}\right>$ ratio for \s{Ar+Sc} collisions an approximation had to be made. In the absence of a measurement of $\langle \pi^+ \rangle$ the mean negative pion multiplicity $\langle \pi^- \rangle$ \cite{naskret} was used instead assuming isospin symmetry: $\langle \pi^+ \rangle \approx \langle \pi^0 \rangle \approx \langle \pi^- \rangle$.

The analysis of particle ratios: $\left<K^{+}\right>/\left<\pi^{+}\right>,~\left<K^{-}\right>/\left<\pi^{-}\right>$ and $\left<K^{+}\right>/\left<K^{-}\right>$ reveals a general trend in particle production properties: measurements obtained for central \s{Ar+Sc} interactions lie between results from \s{p+p} interactions and central \s{Pb+Pb} collisions:

\begin{itemize}\setlength\itemsep{-0.1cm}

	\item  The $\left<K^{+}\right>/\left<\pi^{+}\right>$ ratio (see Fig.~\ref{fig:kpi}) follows the same trend for each beam momentum, namely an increase with system size (quantified by the mean number of wounded nucleons $\langle W \rangle$). \s{Ar+Sc} data points are approaching those of \s{Pb+Pb} collisions towards higher energies. Similar behavior was already observed for the energy dependence of the produced mean pion multiplicity per wounded nucleon~\cite{naskret,naskret2}.

	\item The $\left<K^{-}\right>/\left<\pi^{-}\right>$ ratio (see Fig.~\ref{fig:nkpi}) for \s{Ar+Sc} collisions lies between the values from \s{p+p} and \s{Pb+Pb} interactions. There is a noticeable increase of the ratio with increasing collision energy for all collision systems.

	\item The $\left<K^{+}\right>/\left<K^{-}\right>$ ratio (see Fig.~\ref{fig:kpn}) shows no obvious systematic dependence on the system size. However, there appears to be a decrease with increasing collision energy.

\end{itemize}


\section{Summary}

Preliminary results on charged kaon production were obtained at the CERN SPS in the 5\% most violent \s{Ar+Sc} collisions at three beam momenta (30\textit{A}, 40\textit{A}, 75\textit{A} GeV/\textit{c}). 

Rapidity and transverse momentum spectra of identified $K^{+}$ and $K^{-}$ were measured in the forward rapidity region ($y\in[0.8,2.0]$ at 30\textit{A}, 40\textit{A} GeV/\textit{c} and $y\in[0.6,2.0]$ at 75\textit{A} GeV/\textit{c}) and multiplicities in full phase space were estimated.

Transverse momentum spectra indicate a similarity of \s{Ar+Sc} and \s{Pb+Pb} systems. The extrapolation of the inverse slope parameter to mid-rapidity approaches the values obtained for \s{Pb+Pb} collisions (Fig.~\ref{fig:slope}).

Close similarity of spectra in \s{Ar+Sc} and \s{Pb+Pb} collisions was already observed for negatively charged pions obtained with the "$h^-$" method in a larger acceptance region~\cite{ja}.

Rapidity distributions $\frac{dn}{dy}$ were obtained by integration of $p_T$ spectra. A fit to $\frac{dn}{dy}$ was performed in order to interpolate results in the mid-rapidity region. A strong resemblance of the shape of the distribution was observed for \s{Ar+Sc} and \s{Pb+Pb} systems. Using the fit results, mean charged kaon multiplicities in $4\pi$ phase space $\left<K^{+}\right>$ and $\left<K^{-}\right>$ were derived.

The system size dependence of charged kaon production was studied by comparing ratios of mean multiplicites. 

The $\left<K^{+}\right>/\left<\pi^{+}\right>$ ratios obtained for \s{Ar+Sc} lie between those in \s{p+p} and \s{Pb+Pb} systems, approaching the \s{Pb+Pb} values at higher energies.

The $\left<K^{-}\right>/\left<\pi^{-}\right>$ and $\left<K^{+}\right>/\left<K^{-}\right>$ ratios show no clear dependence on system size.

The collision energy dependence is weak in the measured range for all studied particle ratios. A modest monotonic increase is observed for the $\left<K^{-}\right>/\left<\pi^{-}\right>$ ratio, as well as a monotonic decrease for the measurements of $\left<K^{+}\right>/\left<K^{-}\right>$. The ratio of $\left<K^{+}\right>/\left<\pi^{+}\right>$ in \s{Ar+Sc} collisions shows no indications of a horn structure in the studied limited energy range.

More data are required to understand the particle production dynamics in \s{Ar+Sc} collisions.

In the near future \textit{tof} measurements will supplement the presented data in the mid-rapidity region.

The analysis of \s{Ar+Sc} interactions at three additional beam momenta of 13\textit{A}, 19\textit{A} and 150\textit{A} GeV/\textit{c} is ongoing and will provide a wider range for the study of the energy dependence. This year \s{Xe+La} collisions were recorded by NA61/SHINE. Future analysis will provide more complete information concerning the system size dependence and the onset of the signals of deconfinement.

\vspace{1cm}

\noindent \footnotesize{\textit{Acknowledgments}: This work was partially supported by the National Science Centre, Poland grant 2015/18/M/ST2/00125.}

\end{document}